\documentclass[aps,pra,a4paper,reprint,showpacs,nofootinbib,floatfix,floats]{revtex4-1}
\usepackage{pifont}
\usepackage{color}
\usepackage{amsmath}
\usepackage{amssymb}
\usepackage{amsfonts}
\usepackage{times,txfonts}
\usepackage{bbold}
\usepackage{hyperref}
\hypersetup{
    colorlinks=true,
    citecolor=blue,
    linkcolor=blue,
    urlcolor=blue
}
\usepackage{color}
\usepackage{bm}
\usepackage{graphicx}
\usepackage{float}
\usepackage{epsfig}
\usepackage{verbatim}
\usepackage{epstopdf}
\usepackage{natbib}
\usepackage{appendix}

\providecommand{\openone}{\leavevmode\hbox{\small1\kern-3.8pt\normalsize1}}

\begin{document}
\title{Bell inequality test with entanglement between an atom and a coherent state in a cavity}
\author{Jinwoo Park,$^{1,2}$ Mark Saunders,$^{1,2,3}$ Yong-il Shin,$^{1}$
Kyungwon An,$^{1}$ and Hyunseok Jeong$^{1,2}$}

\affiliation{$^{1}$Department of Physics and Astronomy \&
$^{2}$Center for Macroscopic Quantum Control, Seoul National University, Seoul, 151-742, Korea\\
$^{3}$Department of Education, University of Oxford, 15 Norham Gardens, Oxford, OX2 6PY, United Kingdom}

\date{\today}

\begin{abstract}
We study Bell inequality tests with entanglement between a coherent-state field in a
cavity and a two-level atom.
In order to detect the cavity field for such a test, photon on/off measurements and 
photon number parity measurements, respectively, are investigated. When photon on/off measurements are 
used, at least 50\% of detection efficiency is required
to demonstrate violation of the Bell inequality.
Photon number parity measurements for the cavity field can be 
effectively performed 
using ancillary atoms and an atomic detector, which leads to large degrees of Bell 
violations up to Cirel'son's bound. We also analyze decoherence effects in both field and atomic modes
and discuss conditions required to perform a Bell inequality test free from the locality loophole.

\end{abstract}


\maketitle

\section{Introduction}
\label{Chapter: Introduction}

Einstein, Podolsky, and Rosen (EPR) presented an argument known as the 
EPR paradox \cite{Einstein1935}, which
triggered the debate on quantum mechanics versus local realism.
Bell's theorem \cite{Bell1964}
enables one to perform experiments 
in which failure of local realism is demonstrated by the violation of Bell's 
inequality.
Various versions of Bell's inequality have been developed
including Clauser, Horne, Shimony and Holt (CHSH)'s one \cite{Clauser1969}, 
and substantial amount of experimental efforts have been devoted to the successful
demonstration of violation of Bell's inequality.
So far, many experiments have been performed to show violation of Bell-type
inequalities,
and most physicists now seem to believe that local realism can be violated.

On the other hand, all the experiments performed to date
are subject to some loopholes, so that
the experimental data can still be explained somehow based on a classical (often
impellent) argument.
Experiments using optical fields
\cite{Freedman1972,Aspect1981,Tittel1998,Weihs1998} typically
suffer from the ``detection loophole'' \cite{Pearle1970}, and
recent experiments using atomic states \cite{Rowe2001, Matsukevich2008} 
with the maximum separation of $\sim1$ m \cite{Matsukevich2008}, suffer from the
``locality loophole'' \cite{Bell1981}.
While most of Bell inequality tests have been performed using entangled optical fields
\cite{Freedman1972,Aspect1981,Tittel1998,Weihs1998},
it is an interesting possibility to explore Bell inequality tests using
atom-field entanglement  
\cite{MSKim2000,Milman2005,Simon2003,Volz2006,Brunner2007},
particularly for a loophole-free test.
In fact, there exist theoretical proposals for a loophole-free Bell inequality test using 
hybrid entanglement between atoms and photons \cite{Simon2003,San2011,Spa2011} and
relevant experimental efforts \cite{Moehring2004, Volz2006,Matsukevich2008} have been reported.

In this paper, we study Bell inequality tests with an entangled state of
a two-level atom and a coherent-state field. 
When the amplitude of the coherent state is large enough,
such an entangled state is often called a 
``Schr\"odinger cat state'' (e.g. in Ref.~\cite{Wodkiewicz2000})
as an analogy of Schr\"odinger's paradox where entanglement 
between a microscopic atom
and a classical object is illustrated \cite{Schrodinger1935}.
Entanglement between
atoms and coherent states has been 
experimentally demonstrated using cavities  \cite{Brune1996,Guerlin2007,Deleglise2008}.

In our study, photon on/off measurements and photon number
parity measurements, respectively, are employed in order to detect the cavity field.
We find that when photon on/off measurements are used, at least 50\%
of detection efficiency is required to demonstrate violation of the Bell-CHSH inequality. 
One may effectively perform photon number parity measurements for the cavity field using
ancillary probe atoms and an atomic detector so that
nearly the maximum violation of the Bell-CHSH inequality can be achieved.

The remainder of this paper is organized as follows. 
In Sec.~\ref{sec: Bell Inequality Tests}, we 
briefly discuss the atom-field entanglement under consideration
and review basic elements of Bell inequality tests
in our framework.
We then investigate the Bell-CHSH inequality with photon on/off measurements and 
parity measurements, respectively, 
in Sec.~\ref{Chapter: Bell CHSH inequalities of the Schrodinger cat state}.
Sec.~\ref{Chapter: Bell CHSH inequality with Wigner functions : Practical Scheme with Probing}
is devoted to the investigation of the Bell-CHSH inequality test using indirect measurements 
within a `circular Rydberg atom'-`microwave cavity' system.
In Sec.~\ref{sec:Decoherence and Loopholes},
we analyze decoherence effects in both field and atomic modes.
This analysis enables us to provide quantitative information on the requirements
to perform a loophole-free Bell test.
We conclude with final remarks in Sec.~\ref{sec:con}.

\section{Basic elements for Bell inequality tests}
\label{sec: Bell Inequality Tests}

We are interested in testing the Bell-CHSH inequality with 
an atom-field entangled state:
\begin{equation}
| \Psi \rangle_{AC} = 
\frac{1}{\sqrt{2}}
\left(
|e \rangle_{A}
|\alpha \rangle_{C}
+
|g \rangle_{A}
|-\alpha \rangle_{C}
\right),
\label{eq:bipartite initial state}
\end{equation}
where $|e \rangle_{A}$ ($|g \rangle_{A}$) is the excited (ground) state for the atomic mode $A$,
and $|\pm\alpha \rangle_{C}$ are coherent states of amplitudes $\pm\alpha$ for the field mode $C$.
States (\ref{eq:bipartite initial state}) for reasonably large values of $\alpha$
are considered entanglement between a microscopic system and a classical system
 \cite{Wodkiewicz2000,Jeong2006Sep,Martini2008Jun,Spagnolo2010Nov}.
There have been studies on Bell inequality tests with 
this type of entangled state \cite{Wodkiewicz2000}, and
similar states such as entanglement between an atom and a single photon 
\cite{MSKim2000,Milman2005,Simon2003,Volz2006}
and entanglement between coherent states
\cite{Munro2000,Wilson2002,Jeong2003,JeongSole2008,Jeong2009,Jeong2006Sep,JeongAn2006,Stobinska2007,Gerry2009,LeeJeong2009}.
Experimental demonstration of state (\ref{eq:bipartite initial state}) has been performed 
using a system composed of a circular Rydberg atom and a microwave cavity field
\cite{Brune1996,Guerlin2007,Deleglise2008}.

In order to test a Bell type inequality, a bipartite entangled state
should be shared by two separate parties.
After sharing the entangled state, each of the two parties may locally perform 
appropriate unitary operations and dichotomic measurements.
Violation of the Bell-CHSH inequality can be obtained
by choosing certain values for the parameters of the unitary operations.
The correlation function is defined as the expectation value of the joint measurement 
\begin{equation}
E(\zeta, \beta)= \langle
 \hat E_{A}(\zeta) 
\otimes \hat E_{C}(\beta)  \rangle,
\end{equation}
where $\hat E_{A}(\zeta)=\hat U_{A}^{\dag}(\zeta)\hat\Gamma_{A}\hat U_{A}(\zeta)$
is a dichotomic measurement $\hat \Gamma_{A}$ combined
with unitary operation $\hat U_A (\zeta)$ parameterized by $\zeta$, and
$\hat E_{C}(\beta)$ can be defined accordingly.
The Bell function $\cal{B}$ is then defined as 
\begin{equation}
\begin{split}
{\cal B}
 = \left| E(\zeta, \beta)+E(\zeta', \beta)+E(\zeta, \beta')-E(\zeta', \beta') \right|,
\label{eq:Bell CHSH}
\end{split}
\end{equation}
which should obey the inequality forced by local realism, {\it i.e.},
${\cal B}
\le 2$.
The maximum bound for the absolute value of the Bell function 
is $2\sqrt{2}$, known as Cirel'son's bound \cite{Cirelson1980}.

\label{sec:Dichotomic Measurement in 2 Dimensional Space}

An atomic dichotomic measurement can be represented by a 2 by 2 matrix
\begin{equation}
\hat  \Gamma=
\left( {\begin{array}{*{20}c}
   1 & 0  \\
   0 & { - 1}  \\
\end{array}} \right)
\label{eq:dchm}
\end{equation}
where we choose the basis as \{$|e \rangle$, $|g \rangle$\}.
We define the displaced  dichotomic measurement $\hat \Gamma(\zeta)$
with the atomic displacement operator $\hat{D}(\zeta)$ as
\begin{equation}
\hat \Gamma(\zeta)=\hat{D}(\zeta){\hat{\Gamma}} {{\hat{D}}}^{\dag}(\zeta)
\label{eq:atom A measurement}
\end{equation}
with
\begin{equation}
\begin{split}
&\hat{D}(\zeta)
=
\exp{[{\zeta\hat{\sigma}_{+}-\zeta^{*}\hat{\sigma}_{-}}]}
=
\left(
\begin{array}{cc}
 \cos{\left| \zeta \right|}   &  \frac{\zeta}{\left| \zeta \right|}\sin{\left| \zeta 
\right|}   \\
 -\frac{\zeta^{*}}{\left| \zeta \right|}\sin{\left| \zeta \right|}   &  \cos{\left| 
\zeta \right|}, 
\end{array}
\right), \\
&\zeta(\theta,~\phi)=-\frac{\theta}{2}e^{-i \phi},
\end{split}
\label{eq:2D displacement operator}
\end{equation} 
and $0\le \theta \le \pi$ and $0\le \phi \le 2\pi$, 
where $\hat{\sigma}_{\pm}$ are the standard ladder operators in the $2$-dimensional
Hilbert space.
We note that $\hat{D}(\zeta)$ corresponds to a single qubit rotation for an atomic
qubit and it can be achieved by applying a Ramsey pulse to the atom \cite{Haroche2006}.
We consider  measurement $\hat \Gamma(\zeta)$ for the atomic mode $A$
throughout the paper, 
while some different measurement schemes are considered for the field mode $C$.

\section{Bell-CHSH inequality tests with atom-field entanglement}
\label{Chapter: Bell CHSH inequalities of the Schrodinger cat state}

\subsection{On/off measurement for field mode}

We first investigate the Bell-CHSH inequality with photon on/off
measurements and the displacement operator for the cavity field mode.
The displaced on/off measurement for the field $C$ is
\begin{equation}
\begin{split}
\hat{\cal{O}}_{C}(\beta) 
& = \hat{\cal D}_{C}(\beta)\Big(\sum_{n=1}^{\infty}
|n \rangle \langle n|
-
|0 \rangle \langle 0|\Big)\hat{\cal D}^{\dag}_{C}(\beta), \\
\end{split}
\label{eq:bosonic displaced on/off}
\end{equation}
where $\hat{\cal D}_{C}(\beta) = 
\exp[{\beta}\hat{a}_C^\dagger - {\beta}^*\hat{a}_C]$ is the 
displacement operator with the field annihilation (creation) operator  $\hat{a}_C$ ($\hat{a}_C^\dagger$)
and $\beta$ as the displacement parameter for field $C$.

We model a photodetector with efficiency $\eta$ by
a perfect photodetector together with a beam splitter
of transmissivity  $\sqrt{\eta}$  in front of it \cite{Yuen1980}. The signal field 
$C$ is mixed with the vacuum state $\left| 0 \right\rangle  _v$ at a beam splitter.
The beam splitter operator between modes $C$ and $v$ is  
$\hat{B}_{Cv}=\exp[(\cos^{-1}{\sqrt\eta})
(\hat{a}_C^\dagger \hat{a}_v-\hat{a}_C\hat{a}_v^\dagger)/2]$ \cite{Campos1989},
where $\hat{a}_v$ ($\hat{a}_v^\dagger$) is the field annihilation (creation) operator for the ancilla mode $v$.
After passing through the beam splitter, the atom-field entangled state 
$| \Psi \rangle_{AC}$ is changed to a mixed state as
\begin{equation} 
\begin{split} 
\rho^\eta_{AC}=&{\rm Tr }_v \left[ \hat B_{Cv}
\Big(| \Psi \rangle\langle \Psi|\Big)_{AC} \otimes 
\Big(| 0 \rangle\langle 0|\Big)_v \hat{B}_{Cv}^{\dag} \right] \\ 
=& 
\frac{1}{2} \Big\{ | e \rangle  \langle e | \otimes 
| \sqrt{\eta} \alpha \rangle\langle \sqrt{\eta} \alpha | 
+| g \rangle\langle g | \otimes | -\sqrt{\eta} \alpha \rangle \langle -\sqrt{\eta} \alpha | \\ 
&~~~+e^{ - 2(1 - \eta )\left| \alpha \right|^2 } 
| e \rangle \langle g | \otimes | \sqrt{\eta} \alpha \rangle
 \langle -\sqrt{\eta} \alpha | \\ 
&~~~+e^{ - 2(1 - \eta )\left| \alpha \right|^2 }
 | g \rangle \langle e | \otimes | -\sqrt{\eta} \alpha \rangle
 \langle \sqrt{\eta} \alpha | 
\Big\}_{AC}. 
\label{eq:beamsplitter} 
\end{split} 
\end{equation} 
The correlation function with the photon detection efficiency $\eta$ is the expectation value of 
$\hat{{\Gamma}}_{A} (\zeta) \otimes \hat{\cal{O}}_{C} (\beta)$ for state (\ref{eq:beamsplitter}) as
\begin{equation}
\begin{aligned}
E_{\cal{O}}(\zeta,\beta;\eta)
=&{\rm Tr} \Big[\rho^\eta_{AC}  \hat{{\Gamma}}_{A} (\zeta) \otimes 
\hat{\cal{O}}_{C} (\beta) \Big] \\
=&-e^{-|\beta|^2-|\alpha|^2 \eta -2 |\alpha| |\beta| \sqrt{\eta } \cos\Phi } \cos\frac{\theta }{2}\\
&+e^{-|\beta|^2-|\alpha|^2 \eta +2 |\alpha| |\beta| \sqrt{\eta } \cos\Phi } \cos\frac{\theta }{2}\\
&+e^{-2 |\alpha|^2} \cos\phi  \sin\frac{\theta }{2}\\
&-2 e^{-2 |\alpha|^2-|\beta|^2+|\alpha|^2 \eta } 
\cos\left(\phi -2 |\alpha| |\beta| \sqrt{\eta } \sin\Phi
\right) 
\sin\frac{\theta }{2},\\
\label{eq:qtypecorrelation}
\end{aligned}
\end{equation}
where $\alpha= |\alpha| e^{i \Phi_{\alpha}}$,
$\beta = |\beta| e^{i \Phi_{\beta}}$,
and $\Phi= \Phi_{\beta} - \Phi_{\alpha}$
with real phase parameters $\Phi_{\alpha}$ and $\Phi_{\beta}$.
The Bell function is immediately obtained using Eqs.~(\ref{eq:Bell CHSH}) and
(\ref{eq:qtypecorrelation}).

\begin{figure}[t]
\begin{centering}
\epsfig{file=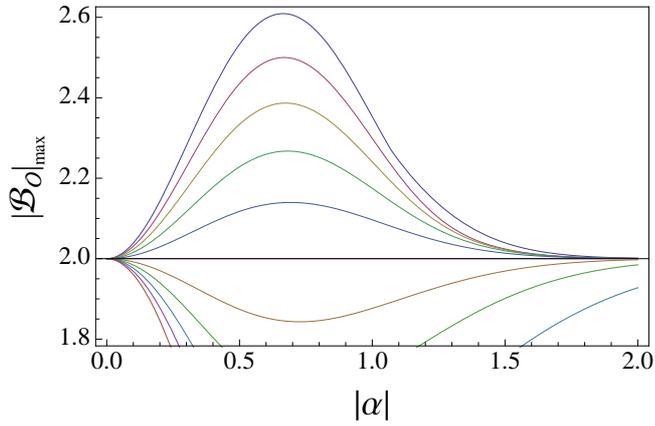,width=8.5cm}
\caption{(Color online)
Numerically optimized values of Bell functions
$\cal{B}_{\cal{O}}$ with displaced on/off measurements 
against amplitude $\alpha$ of state (\ref{eq:bipartite initial state}).
The detection efficiency ranges in value from $\eta = 0$  
(lower curve) to $\eta=1$ (upper curve), with intervals of 0.1 shown by the family of curves.
The horizontal line corresponds to the case of $\eta=0.5$,
which coincides with the classical limit of the Bell-CHSH inequality.
}
\label{fig:B_alpha_qtype}
\end{centering}
\end{figure}

\begin{figure}[t]
\begin{centering}
\epsfig{file=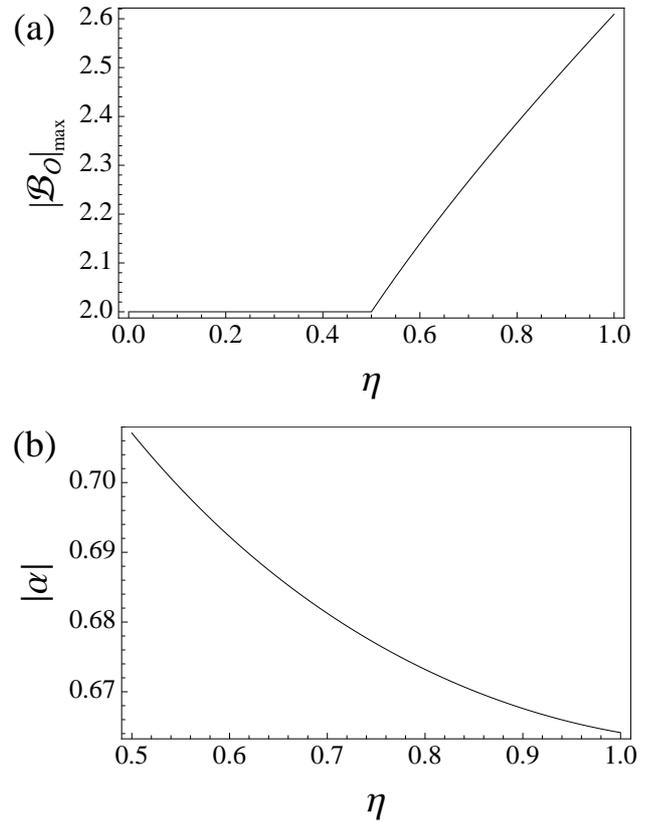,width=8.5cm}
\caption{
(a) Numerically optimized values of Bell function $\cal{B}_{\cal{O}}$
with displaced on/off measurements against detection efficiency $\eta$.
The local realistic bound, 2, is violated for $\eta\ge0.5$.
(b) Plot of optimizing values  of $|\alpha|$ with respect to $\eta$.
}
\label{fig:B_eta_qtype}
\end{centering}
\end{figure}

Using the method of steepest descent \cite{Pres1988},
we numerically find optimized values,
$|{\cal B}_{\cal{O}}|_{\rm max}$, {\it i.e.}, 
absolute values of the Bell function maximized over variables
$\zeta$, $\zeta'$, $\beta$ and $\beta'$.
We plot the results against amplitude $|\alpha|$
for various choices of the detection efficiency from $\eta=0$ to 
$\eta=1$ (from bottom to top), where $\eta$ differs by $0.1$ between closest curves
in Fig.~\ref{fig:B_alpha_qtype}. 
Assuming a real positive value of $\alpha$, we find that
the optimizing conditions
can also be obtained as
\begin{equation}
\zeta=\frac{\pi}{2},~~ \zeta'=0, ~~\beta=-\beta'=|\beta|,
\label{eq:BOopt}
\end{equation}
where $|\beta|$ satisfies
\begin{equation}
2 |\beta| e^{2(\eta -1){|\alpha|}^2}=e^{-2{|\alpha|}
|\beta|\sqrt{\eta }} \left(|\beta|+{|\alpha|} \sqrt{\eta }\right)
- e^{2{|\alpha|} |\beta|\sqrt{\eta }} \left(|\beta|-{|\alpha|}
\sqrt{\eta }\right).
\end{equation}
As expected, the perfect detection efficiency, $\eta=1$, gives
the higher violation up to $|{\cal B}_{\cal O}|_{\rm max}\approx2.61$ when $|\alpha|\approx0.664$.
A Bell violation of $|{\cal B}_{\cal O}|_{\rm max}\approx 2.39$ ($|{\cal B}_{\cal O}|_{\rm max}\approx2.14$) is obtained 
for $\eta=0.8$ ($\eta=0.6$) when $|\alpha|\approx0.673$ ($|\alpha|\approx0.692$).

When $|\alpha|=0$, no violation occurs
because state (\ref{eq:bipartite initial state}) contains no entanglement.
As $|\alpha|$ increases, the Bell violation becomes higher until $|\alpha|\sim 0.7$. 
However, as shown in Fig.~\ref{fig:B_alpha_qtype}, as $|\alpha|$ keeps increasing, 
the degree of the Bell violation decreases towards zero 
even though the state has larger entanglement.
This result is due to the fact that when $|\alpha|$ is large, 
the probability of detecting the vacuum for the field mode diminishes.
Obviously, if photon on/off detection excludes one of the two possible results,
violation of the Bell-CHSH inequality will not occur regardless of the degree of entanglement.
This is in agreement with a previous result in Ref.~\cite{Jeong2003}
where the Bell-CHSH inequality with entangled coherent states, $|\alpha\rangle|-\alpha\rangle
-|-\alpha\rangle|\alpha\rangle$ (without normalization), was considered with on/off detection.

It should be noted that in Fig.~\ref{fig:B_alpha_qtype},
the Bell functions for $\eta= 0.5$ overlaps with the horizontal line that indicates
the classical limit $2$. In fact, the photon detector efficiency
should be higher than 0.5 in order to see a Bell violation as shown in Fig.~\ref{fig:B_eta_qtype}(a).
Figure~\ref{fig:B_eta_qtype}(b) shows that the optimizing values of $|\alpha|$ are 
within the range of $0.66<|\alpha|<0.71$  for any of $\eta\geq 0.5$.
We also note a previous result \cite{Brunner2007} that efficiency of $0.43$ can be tolerated if
a different type of Bell inequality \cite{Gisin1999} is used with a nonmaximally
entangled state and a perfect atomic measurement.

\subsection{Photon number parity measurement for field mode}

\begin{figure}[t]
\begin{centering}
\epsfig{file=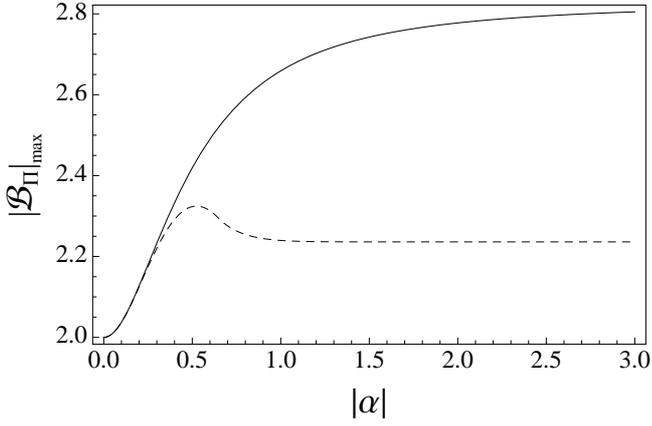,width=8.5cm}
\caption{Numerically optimized values of Bell function ${\cal B}_{\Pi}$
with displaced parity measurements
against $|\alpha|$.
The solid curve corresponds to the absolute 
values of the Bell function maximized  over arbitrary $\zeta$, $\zeta'$, $\beta$ and $\beta'$, while
the dashed curve corresponds to those values  maximized  over arbitrary $\beta$ and $\beta'$,
but real $\zeta$ and $\zeta'$.
   }
\label{fig:B_alpha_wtype}
\end{centering}
\end{figure}

We now consider the displaced photon number parity measurement for the field mode 
\begin{equation}
\begin{split}
\hat{\Pi}_{C}(\beta) 
& = \hat{\cal D}_{C}(\beta)
\Big(\sum_{n=0}^{\infty}{
|2n \rangle \langle 2n|
-
|2n+1 \rangle \langle 2n+1|}\Big)
\hat{\cal D}^{\dag}_{C}(\beta). 
\end{split}
\label{eq:bosonic displaced parity}
\end{equation}
Using Eq.~(\ref{eq:bipartite initial state}) and the measurement operators defined above,
it is straightforward to get 
\begin{equation}
\begin{split}
E_{\Pi}(\zeta, \beta)=&\left\langle \hat{\Gamma}_{A}
 (\zeta) \otimes \hat{\Pi}_{C} (\beta) \right\rangle\\
=&e^{ - 2{|\beta|}^2 } \sin \theta \cos [ 4|\alpha| |\beta|\sin \Phi -\phi  ] \\
&+ e^{ - 2({|\alpha|}^2  + {|\beta|}^2 )} \cos \theta \sinh [4|\alpha| |\beta|\cos \Phi ],
\end{split}
\label{eq:wignertypecorrelation}
\end{equation}
and the corresponding Bell function, ${\cal{B}}_{\Pi}$.
We present the numerically optimized Bell function, $|{\cal{B}}_{\Pi}|_{\rm max}$,
 against $|\alpha|$ in Fig.~\ref{fig:B_alpha_wtype},
where Bell violation occurs for any nonzero $\alpha$.
Note that the atomic displacement operator corresponds to
a single-qubit rotation for the atomic mode.
It was argued that
the field displacement plays a similar role to
approximately rotate a coherent-state qubit \cite{Jeong2003}.
If we restrict the atomic displacement parameters ($\zeta$ and $\zeta'$) to be real,
our test becomes identical to the one in Ref.~\cite{Wodkiewicz2000} and the result corresponds 
to the dashed curve in Fig.~\ref{fig:B_alpha_wtype}. 
However, it is not sufficient to reveal the maximal violation of the atom-field entangled state
(\ref{eq:bipartite initial state}). In our numerical analysis,
${\cal{B}}_{\Pi}$
is optimized with respect to complex $\zeta$, $\zeta'$, $\beta$, and $\beta'$
that results in the solid curve in Fig.~\ref{fig:B_alpha_wtype}.
Assuming that $\alpha$ is a real positive value, 
the optimizing conditions for ${\cal B}_\Pi$ are found as
\begin{equation}
\zeta=-\pi/4, ~~\zeta'=i\pi/4,~~ 
\beta=-\beta^\prime=i|\beta|,
\label{eq:optparity}
\end{equation}
where $|\beta|$ satisfies
\begin{equation}
(|\alpha|-|\beta|)/(|\alpha|+|\beta|)=\tan{4 |\alpha| |\beta|}
\label{eq:optparitybeta}
\end{equation}
{\it and} is nearest to zero.
As amplitude $|\alpha|$ increases, the degree of Bell violation rapidly gets larger 
up to Cirel'son's bound $2\sqrt{2}$.

\section{Approach using indirect measurement}
\label{Chapter: Bell CHSH inequality with Wigner functions : Practical Scheme with
Probing}

\begin{figure}[t]
\begin{centering}
\epsfig{file=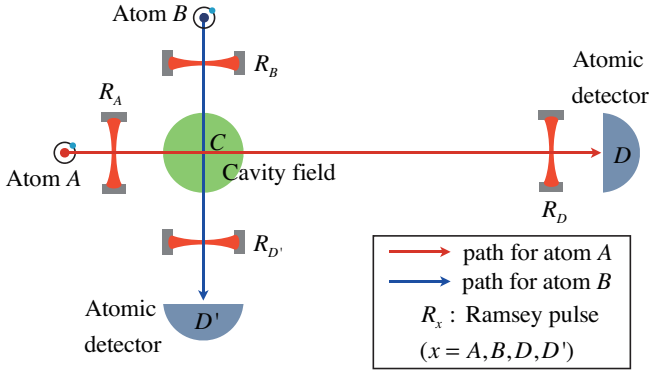,width=8.5cm}
\caption{(Color online) Schematic of the proposal. The horizontal arrow 
is to describe
the entangled state (\ref{eq:bipartite initial state}) generation
(with $R_A$ and $C$) and measurement for atom $A$.
The vertical arrow depicts  the indirect parity measurement of the cavity field using 
ancillary atom $B$. 
}
\label{fig:probe}
\end{centering}
\end{figure}

In this section, we discuss physical implementations of the Bell-CHSH inequality test using
displaced parity measurements in a `circular Rydberg atom'-`microwave cavity' 
configuration. 
Generation schemes for atom-field entangled states (\ref{eq:bipartite initial state})
have been theoretically studied and experimentally implemented
\cite{Brune1992, Davidovich1996,  Haroche2006, Raimond2001}. 
In the case of a scheme based on the off-resonant interaction \cite{Haroche2006},
the required interaction Hamiltonian is
\begin{equation}
\hat{H}_{I}=\hbar \chi [ (\hat a^\dag  \hat a + 1) \left| e \right\rangle 
\left\langle e \right| - \hat a^\dag  \hat a\left| g \right\rangle \left\langle g 
\right| ],
\label{eq:effective Hamiltonian}
\end{equation}
and $\chi=\Omega^2 / (4\delta)$ is the coupling constant 
determined by the vacuum Rabi frequency $\Omega$ and detuning $\delta$ \cite{Haroche2006}.
As shown in Fig.~\ref{fig:probe}, 
$\pi/2$ Ramsey pulse with phase $-\pi/2$ ($R_{A}$)
is applied to
a circular Rydberg atom ($A$) prepared in the excited state $\left| e \right\rangle_A$
\cite{Nussenzveig1993}, 
which results in an atomic superposition state: 
$\left| \phi_{-i}\right\rangle_A  =  \left(
\left| e \right\rangle_A  -i \left| g \right\rangle_A
\right)/{\sqrt 2 }$.
Then, a strong dispersive interaction in 
Eq.~(\ref{eq:effective Hamiltonian})
 between atom $A$ and the cavity field produces
the atom-field entangled state (\ref{eq:bipartite initial state}) for
interaction time $t=\pi/(2\chi)$ \cite{Haroche2006}.

\begin{figure}[t] 
\begin{centering} 
\epsfig{file=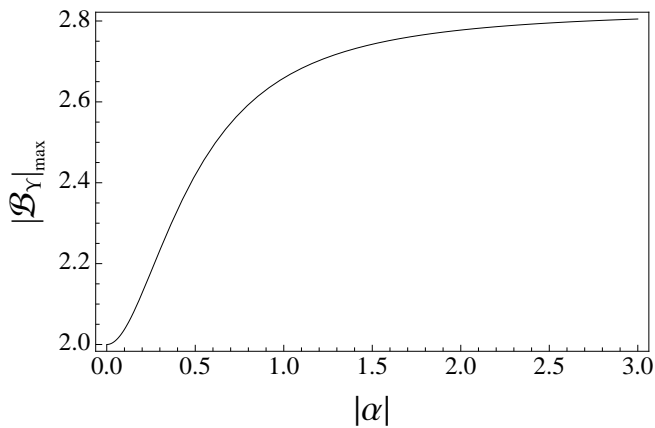,width=8.5cm} 
\caption{ 
Numerically optimized values of Bell function ${\cal B}_{\Upsilon}$
with indirect measurements against $|\alpha|$. 
The result is found to be identical to the one using direct parity measurements 
shown as the solid curve in Fig.~\ref{fig:B_alpha_wtype}. } 
\label{fig:upsilon} 
\end{centering} 
\end{figure}

Direct measurements of the light field in the microwave cavity are difficult to achieve,
while indirect methods for parity measurements of the cavity field may be more feasible 
\cite{Englert1993, MSKim2000, Haroche2006,Deleglise2008}.
A circular Rydberg atom ($B$)
in Fig.~\ref{fig:probe} initially prepared in state $\left| e \right\rangle_B$ 
evolves to a superposition state
$\left| \phi_{-i} \right\rangle_{B}$
by $\pi/2$ Ramsey pulse with phase $-\pi/2$
($R_{B}$), and 
the total state is $|\Psi_{tot}\rangle_{ABC}=
\left| {\Psi } \right\rangle_{AC}  
\left| \phi_{-i} \right\rangle_{B}$.
The displacement operation, 
$\hat {\cal D}^\dagger _C (\beta)=\hat {\cal D}_C (-\beta)$, is then applied to the 
field right before atom $B$ enters the cavity, and
the same type of interaction as Eq.~(\ref{eq:effective Hamiltonian})
between modes $B$ and $C$ follows.
One may indirectly detect the cavity field 
by appropriately choosing the interaction time $t=\pi/(2\chi)$ between
atom $B$ and the field before detecting the atom. 
The interaction time may be controlled by selecting the velocity of atom $B$.
The final measurement for atom $A$, represented by $\hat{\Gamma}_A(-{\pi}e^{-i\phi}/4)$, is performed
using $\pi/2$ Ramsey pulse of phase $\pi-\phi$ ($R_D$) and atomic detector $D$. 
The  measurement on atom $B$,
 {\it i.e.},  $\hat{\Gamma}_B(-\pi/4)$, for indirect probing
is performed 
with the help of $\pi/2$ Ramsey pulse with $\pi$ phase ($R_{D'}$) and atomic detector $D'$.
The measurement operator is then represented as
\begin{equation}
\hat{\Upsilon}_{ B,C} (\beta, t) ={\hat {\cal U}_{  B,C}(\beta,t)}^\dagger \hat {\cal O}_{ B,C}\hat
 {\cal U}_{B,C}(\beta,t),
\label{eq:im}
\end{equation}
where 
\begin{equation*}
\begin{split}
&\hat {\cal O}_{ B,C} =  [\left|  +  \right\rangle \left\langle  +  \right| - \left|  -  
\right\rangle \left\langle  -  \right|]_{B} \otimes {\openone}_{C}, \\ 
&\hat {\cal U}_{ B,C} = e^{ - i\hat{H}_{I} t /\hbar }_{ B,C}\hat {\cal D}_{C}^{\dag}(\beta),
\end{split} 
\end{equation*}
and $\left| \pm \right>=(|e\rangle\pm|g\rangle)/\sqrt{2}$.
The correlation function $E(\zeta,\beta, 
t)=\left\langle \hat{{\Gamma}}_{ A} (\zeta) \otimes
\hat{\Upsilon}_{B,C} (\beta, t) \right\rangle$ is calculated using 
state $|\Psi_{tot}\rangle_{ABC}$ as
\begin{widetext}
\begin{equation}
\begin{split}
E(\zeta,~\beta,~t)=&
  \frac{1}{2}\cos{\theta} e^{(|\alpha|^2  + |\beta|^2 
  - 2|\alpha||\beta|\cos \Phi )( - 1 + \cos 2\chi t)}
  \cos [(|\alpha|^2  + |\beta|^2  - 2|\alpha||\beta|\cos \Phi )\sin 2\chi t]\\
&- \frac{1}
{2}\cos \theta e^{(|\alpha|^2  + |\beta|^2  + 2|\alpha||\beta|\cos \Phi )( - 1 + \cos 2\chi t)} 
\cos [(|\alpha|^2  + |\beta|^2  + 2|\alpha||\beta|\cos \Phi )\sin 2\chi t]    \\
 & +\frac{1}
{2}\sin \theta e^{ - |\alpha|^2  - |\beta|^2  - (|\alpha|^2  
- |\beta|^2 )\cos 2\chi t - 2|\alpha||\beta|\sin 
\Phi \sin 
2\chi t} 
\cos [\phi  - (|\alpha|^2  - |\beta|^2 )\sin 
2\chi t + 2|\alpha||\beta|\sin \Phi ( - 1 + \cos 2\chi t)]   \\
& + \frac{1}
{2}\sin \theta e^{ - |\alpha|^2  - |\beta|^2  
- (|\alpha|^2  - |\beta|^2 )\cos 2\chi t + 2|\alpha||\beta|\sin 
\Phi \sin 
2\chi 
t} \cos [\phi  + (|\alpha|^2  - |\beta|^2 )\sin
2\chi t + 2|\alpha||\beta|\sin \Phi ( - 1 + \cos 2\chi t)]
\end{split}
\label{eq:correlation}
\end{equation}
\end{widetext}
and the Bell function, ${\cal B}_\Upsilon$, is accordingly obtained. 
As expected, 
the optimizing conditions for $|{\cal B}_\Upsilon|_{\rm max}$
are identical to those for $|{\cal B}_{\cal O}|_{\rm max}$ in Eqs.~(\ref{eq:optparity}) and
(\ref{eq:optparitybeta}) with an additional condition, $t=t'=\pi/2\chi$.
Our numerical study confirms that the optimized Bell function $|{\cal B}_\Upsilon|_{\rm max}$
plotted with the abovementioned conditions in Fig.~\ref{fig:upsilon}
exactly overlaps with the solid curve 
in Fig.~\ref{fig:B_alpha_wtype} as shown .
This result is due to the fact that the indirect measurement (\ref{eq:im})
is basically equivalent to the displaced parity measurement 
(\ref{eq:bosonic displaced parity}) on the cavity field
when $t$ is chosen to be $\pi/(2\chi)$ \cite{Englert1993}.
{\it I.e.}, the measurement on 
atom $B$ in the basis $\{ \left| + \right>,~\left|-\right> \}$
after the interaction time $t = \pi/(2\chi)$ is equivalent to the 
parity measurement on the cavity-field.
In fact, it can be shown that the correlation functions (\ref{eq:correlation})
with $t =\pi/(2\chi)$ and 
(\ref{eq:wignertypecorrelation}) are identical. Of course, if we restrict $\zeta$
to be real, the optimized plot of the Bell function $|{\cal B}_\Upsilon|_{\rm max}$
approaches the dashed curve in  Fig.~\ref{fig:B_alpha_wtype}.

\section{Decoherence and Loopholes}
\label{sec:Decoherence and Loopholes}

It is not difficult to predict that
decoherence effects due to the cavity-field dissipation and the spontaneous emission of the atoms
will obstruct Bell violations.
This is particularly important when one intends to demonstrate
a Bell violation free from the loopholes.
In this section, we consider decoherence effects with realistic 
conditions for the Bell-CHSH inequality test using parity measurements
and suggest quantitative requirements 
to perform a loophole-free Bell test.

\subsection{Decoherence effects in the cavity-atom system}

\begin{figure}[t]
\begin{centering}
\epsfig{file=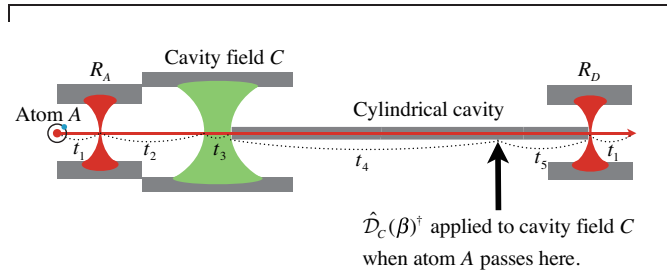,width=8.5cm}
\caption{(Color online) Sideview of the atom $A$'s path with intervals of time. Each interval denotes an amount of time required for atom $A$ to pass through the region related with atomic velocity $v$. We note
that the distance $l=v \times (t_4+t_5)$, which corresponds to the length of the cylindrical cavity, is a crucial factor in a loohole-free Bell inequality test.}
\label{fig:tablefigure}
\end{centering}
\end{figure}

There are two main effects that cause decoherence  in our Bell inequality test, {\it i.e.}, spontaneous emissions
from atoms and cavity field dissipations.  
In the atom-cavity system under consideration, one (or both) of
these two effects may occur. 
The master equation which determines the time-evolution of the density operator, $\hat {\rho}(t)$, under the atom-field interaction with spontaneous emissions and cavity dissipations is
\begin{equation}
\frac{{d\hat \rho (t)}}{{dt}} = \frac{1}{{i\hbar }}[\hat H_{I} ,\hat \rho (t)] + 
{\cal L}\hat \rho (t),
\label{eq:mi}
\end{equation}
with the Linblad decohering term $\cal L$ defined as
\begin{equation}
\begin{split}
{\cal L}\hat \rho  \equiv &\kappa (2\hat a\hat \rho \hat a^\dag   - \hat 
a^\dag  \hat a\hat \rho  - \hat \rho \hat a^\dag  \hat a)\\
&+{\gamma}(2\hat
\sigma_{-} \hat \rho \hat \sigma_{+}   - \hat \sigma_{+}  \hat \sigma_{-}\hat \rho  - 
\hat \rho \hat \sigma_{+}  \hat \sigma_{-}),
\label{eq:Ld}
\end{split}
 \end{equation}
where $\kappa$ is the dissipation rate of cavity field, and $\gamma$ is the spontaneous emission rate.

It is known that the spontaneous emission rate of an atom
can be significantly reduced by engineering the shape of the cavity that
contains the atom  \cite{Hulet1985,Kakazu1996}. 
A complete inhibition of spontaneous emission 
was suggested using a cylindrical metal cavity
with a diameter shorter than
${1.8412 c }/{\omega_{0}}$, 
where $\omega_{0}$ is the 
transition rate between atomic states $|e\rangle$ and $| g \rangle$
and $c$ is the speed of light
     \cite{Kakazu1996}. 
For our setup,
the transition rate can be
taken from Ref.~\cite{Deleglise2008} as
$\omega_{0}=51.1 \rm{GHz}$. This means that the diameter should be smaller than $3.44 \rm{mm}$
that is experimentally achievable.
As seen in Fig.~\ref{fig:tablefigure}, a long cylindrical cavity may be used
between cavity $C$ and Ramsey zone $R_D$ to inhibit spontaneous emission.

The spontaneous emission rate $\gamma_c$ inside the cavity $C$ in Fig.~\ref{fig:probe}
is also generally different from the spontaneous emission rate  $\gamma_0$ in the vacuum.
It is known that $\gamma_{c}$ can be 
calculated by approximating the cavity in the one 
dimension while considering the effect of the atomic motion as described in Ref.~\cite{Wilkens1992}.
In our case, $\gamma_{c}=\rm{4.08~Hz}$ is obtained based on the result of 
Ref.~\cite{Wilkens1992} from the spontaneous emission rate in the vacuum, 
$\gamma_0=1/(2T_{0})$ ($T_{0}=\rm{36~ms}$ is the atomic life time in the vacuum \cite{Haroche2006}) and related realistic parameters in a recent experiment~\cite{Deleglise2008}.

\begin{figure}[t]
\begin{centering}
\epsfig{file=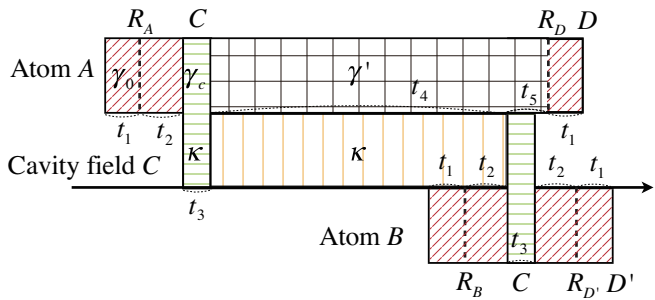,width=8.5cm}
\caption{(Color online)
A timeline for decoherence with dynamical parameters related in each regions (from left to right).
The top line is for atom $A$, the middle for cavity field $C$, and the bottom for atom $B$.
The times when Ramsey pulses are applied are described as vertical dashed lines.
We consider a Ramsey pulse application as an instant event as a Ramsey pulse lasts as short as 
 $\rm{1~\mu s}$ order \cite{Raimond2001}. 
Regions are differently hatched depending on the types of dynamics.
In the diagonally hatched regions, atoms $A$ and $B$ travel in free spaces with the spontaneous emission rate $\gamma_0$
before and after Ramsey pulses as shown in Fig.~\ref{fig:probe}.
In the cross-hatched region, atom $A$ travels in a cylindrical cavity with the inhibited
spontaneous emission rate $\gamma^\prime$.
In the vertically hatched region, the cavity dissipation with rate $\kappa$ occurs in the cavity ($C$) field.
The horizontally hatched regions correspond to the dynamics of the atom-field interaction $\hat{H}_I$ 
in the main cavity $C$ together with spontaneous emission $\gamma_c$ and cavity dissipation $\kappa$.  
Abbreviations $C$, $D$, $D'$, $R_A$, $R_B$, $R_D$, and $R_{D'}$
are consistent with those in Fig.~\ref{fig:probe}.} 
\label{fig:timetable}
\end{centering}
\end{figure}

Considering the discussions above, we present a timeline of decoherence effects in Fig.~\ref{fig:timetable}
together with 
time intervals required to pass through certain parts of the apparatus as follows (also depicted in Fig.~\ref{fig:tablefigure}):~
 $t_1$ is a half of the time required for an atom to pass through a cavity used for Ramsey pulse application,
$t_2$
is a half of the time required for an atom to pass through Ramsey pulse and the main cavity (C) without cavity waist,
$t_3$ for an atom to pass through the main cavity(C)'s waist ($\pi/2\chi$),
$t_4$ for atom $A$ to pass through the long cylindrical cavity before the field displacement operation on the cavity field,
$t_5$ for atom $A$ to pass through the remainder of the long cavity after the field displacement operation, and
$t_6$ for atomic detection at $D$ or $D'$.

Let us first consider the pathway of atom A, which corresponds to the top line of Fig.~\ref{fig:timetable}.
Atom $A$ undergoes spontaneous emission before and after the Ramsey pulse $R_A$ with rate $\gamma_0$ (diagonally hatched part).
Atom $A$ then interacts with the cavity field with dissipation rate $\kappa$
under spontaneous emission ($\gamma_c$), which is represented by the horizontally hatched part. After the atom-field interaction, atom $A$ passes through the cylindrical cavity experiencing inhibited spontaneous emission ($\gamma^\prime$). Finally, atom $A$ comes out of Ramsey pulse $R_D$ experiencing spontaneous emission ($\gamma_0$), and is registered at detector $D$.
In the mean while, cavity field $C$ which have interacted with atom $A$ undergoes field dissipation ($\kappa$) while atom $A$ is passing through cylindrical cavity. Then, cavity field $C$ begins to interacts with atom $B$ under spontaneous emission  ($\gamma_c$) and field dissipation ($\kappa$) after displacement operation on it. Atom $B$, used for an indirect measurement, 
experiences spontaneous emission ($\gamma_0$) around Ramsey pulse $R_B$, interaction
with the cavity field ($C$) with spontaneous emission ($\gamma_c$), and 
spontaneous emission ($\gamma_0$) before detection $D^\prime$.

Here, we take the photon storage time
$T_{C}=\rm{0.13 ~s}$ ($\kappa=1/(2T_C)$), $\Omega  = 2\pi  \cdot \rm{49~kHz}$ and $\delta=2\pi  \cdot 
\rm{65~kHz}$ ($\chi=\Omega^2/(4\delta)\approx58{\rm kHz}$)
from recent experiments  \cite{Deleglise2008}. The solution of the master equation for the cavity 
dissipation alone with $H_I$, was examined in Ref.~\cite{Faria1999May}.
In Appendix, we obtain the solution of Eq.~(\ref{eq:mi})
and find an explicit form of the density operator and the correlation function.
The Bell function can be constructed using the correlation function in Eq.~(\ref{eq:correlationfunction})
of Appendix.
Note that we have assumed perfect Ramsey pulses during the procedures.  
Considering cavity dissipation, we employ the same optimizing conditions (\ref{eq:optparity}) except that
$|\beta|$ is chosen to be the values that satisfy 
\begin{equation}
\frac{|\alpha| e^{-\kappa(t_4 + t_3)}-|\beta|}{|\alpha| 
e^{-\kappa(t_4 + t_3)}+|\beta|}=\tan({4 |\alpha| e^{-\kappa(t_4 + t_3)} |\beta|}),
\label{eq:optconbetamod}
\end{equation}
{\it and} is nearest to zero.

\subsection{Bell violation and separations under practical conditions without 
a cylindrical cavity}
\label{subsec:nocylinder}
\begin{figure}[h]
\begin{centering}
\epsfig{file=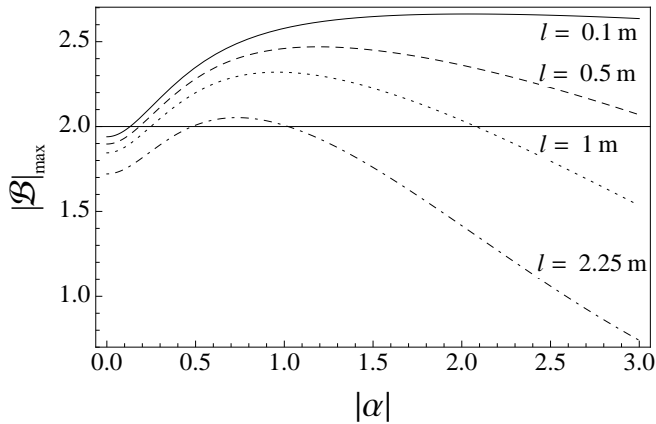,width=8.5cm}
\caption{
The Bell function under realistic conditions discussed in Sec.~\ref{subsec:nocylinder}
are plotted with optimizing conditions in Eqs.~(\ref{eq:optparity}) and (\ref{eq:optconbetamod})
for several different cases of separation $l$.
 As the separation $l$ gets larger, the maximum values of the Bell function decrease.
 The decoherence effects become heavier as $|\alpha|$ gets larger.}
\label{fig:practical}
\end{centering}
\end{figure}

Let us first consider Bell violation depending on the separation $l = v \times (t_4+t_5)$
between both parties {\it without} using a cylindrical cavity (thus $\gamma'=\gamma_0$).
We choose some practical time-interval parameters as $t_1=\rm{80.0 ~\mu s}$, 
$t_2=\rm{166.5 ~\mu s}$,  $t_3=\rm{27.1 ~\mu s}$, $t_6=\rm{20 ~\mu s}$ 
and velocity of an atom $v=\rm{250~m/s}$ \cite{Deleglise2008,Zhou2009P}. 
The Bell function with several choices of $l$ are plotted in Fig.~\ref{fig:practical}. 
The Bell function approaches 
the value near 2.7 when $l=0.1$(meter), but 
it decreases as $l$ gets larger.
Clear Bell violations appear for $l\lesssim 2$(meter), however, this is insufficient for a space-like separation as we shall discuss in the next subsection.

\subsection{Requirements for a Bell test free from the locality loophole with
a cylindrical cavity}
In principle, a Bell test free from the locality loophole 
can be performed using 
a long cylindrical cavity with a low spontaneous emission rate ($\gamma'$) and the main cavity 
with a low dissipation rate ($\kappa$).
In order to close the locality loophole, 
the measurement event for atom $A$ should not affect the measurement event for the cavity field $C$, 
and {\it vice versa} 
\cite{Bell1981}. In other words, the measurement event for atom $A$ should be outside of
the ``back light cone'' from the detection event $D^{\prime}$ in Fig.~\ref{fig:probe}. 
In the same manner, the measurement event for the cavity field $C$ 
should not be in the back light cone from the detection event $D$.
For simplicity, let us first suppose that each measurement process takes place at a single location ($D$ and $D^{\prime}$). 
In our Bell test, 
the time  $t_A$ required to measure atom $A$ is smaller than the time required to measure field $C$ ($t_C$)
due to the indirect measurement scheme for field $C$.
We assume that the measurement event for the field $C$ precedes to the measurement event for atom $A$ by $T$ (the opposite case will require a longer separation between the two parties). 
Then the conditions required to close the locality loophole are
\begin{equation} 
\begin{split} 
d \ge c (T+t_A), \\ 
d \ge c (t_C-T), 
\end{split} 
\label{eq:nonlocalitysimple} 
\end{equation} 
where $d$ is the distance between $D$ and $D^{\prime}$ and 
$c$ is the speed of light.

In order to apply the locality-loophole-free conditions (\ref{eq:nonlocalitysimple})
to our Bell test setup in a more rigorous manner, one needs to
consider locations of the local measurement elements.
In Fig.~\ref{fig:timetable}, one can find that 
the measurement time for atom $A$ ($t_A$) consists of the times for $R_D$ ($t_1$) and $D$ ($t_6$) 
and that for the field ($t_C$) consists of the times for $C$ ($t_3$), $R_D^\prime$ ($t_2+t_1$), and $D^\prime$ ($t_6$). 
A measurement event for each party actually does not take place at a single location, and both
of the measurements are not even on a straight line. Therefore the distance $d$ in Eqs.~(\ref{eq:nonlocalitysimple})
needs to be replaced with the distances from the final detector of one party to the location where the measurement
of the other party begins. 
A careful consideration leads to the conclusion that the following inequalities should be satisfied:
\begin{equation}
\begin{split}
& v (t_3 /2+ t_4 + t_5+t_1 +t_6)\ge c (t_5+t_1+t_6),\\
& v \sqrt{(t_3 /2+ t_2 + t_1+t_6)^2+(t_3 /2+ t_4 + t_5)^2} \\
&~~~~~~~~~~\ge c (t_3+t_2+t_1+t_6-t_5).
\end{split}
\label{eq:nonlocality}
\end{equation}
Using the feasible values of  $t_1$, $t_2$, $t_3$, $t_6$ and $v$ in the previous subsection,
we find the minimum values $t_4=236.0~s$ and $t_5= 96.8 ~\rm{\mu s}$ with which the equalities
hold for Eqs.~(\ref{eq:nonlocality}). Then,
the minimum distance required for a Bell test free from the locality loophole
is found to be $l=\rm{52.99~km}$ \cite{Deleglise2008}.

\begin{figure}[!]
\begin{centering}
\epsfig{file=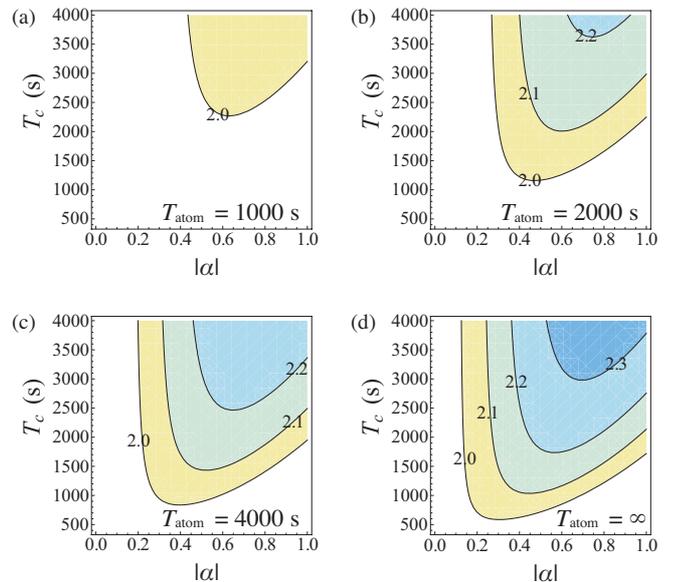,width=8.5cm}
\caption{(Color online) Contour plots of the Bell function with respect to photon
storage time $T_C $ in the main cavity and amplitude $|\alpha|$ of the entangled state. The atomic life time in cylindrical cavity $T_{\rm atom}$ is fixed at 1000, 2000, 4000 and $\infty$ (seconds).
The minimum distance condition $l=52.99$ (km) for a loophole-free Bell test was assumed.  Higher inhibition of spontaneous emission in the cylindrical cavity reduces the required photon storage time in the main cavity. 
}
\label{fig:ideal}
\end{centering}
\end{figure}

We finally consider conditions of the atomic life time $T_{\rm{atom}}=1/(2\gamma^\prime)$
and the photon storage time $T_c=1/(2\kappa)$ required for a Bell test free from the locality loophole.
In Fig.~\ref{fig:ideal}, we
plot the Bell function constructed using Eq.~(\ref{eq:correlationfunction}) in Appendix
with respect to the photon storage time in the main cavity and amplitude $|\alpha|$ of the atom-field entanglement. Here, the extended lifetime of the 
atom in the cylindrical cavity was assumed to be $T_{\rm atom}=$1000, 2000, 4000, and $\infty$ (seconds).
The distance $l$ was assumed to be the minimum distance required for 
a loophole-free Bell test ($52.99$ km).
For example, when $T_{\rm atom}=2000$ (seconds), the photon storage time $T_C\sim 1160$ (seconds) at $|\alpha|\sim 0.47$ is required to see a Bell violation.
If complete inhibition of the spontaneous emission in the cylindrical cavity is possible, ({\it i.e.}, $T_{\rm atom}=\infty$), $T_C\sim 590$ at $|\alpha|\sim0.3$ is required.
Obviously, the stronger inhibition of the spontaneous emission in the cylindrical cavity relaxes the requirement
of the photon storage time in the main cavity to see  Bell violations. 
However, it still requires at least a few hundreds of seconds for
the photon storage time to demonstrate a loophole-free Bell violation,
 while it is only about $0.13 ~{\rm{s}}$ at present \cite{Kuhr2007}.   
It would also be extremely challenging to build a long cylindrical cavity 
that strongly inhibits the spontaneous emission of atom $A$ during such a long life time.

\section{Remarks}
\label{sec:con}

We have investigated Bell-CHSH inequality tests with entanglement between 
a two-level atom and a coherent-state field in a cavity.
In order to detect the cavity field for these tests, photon on/off measurements and 
photon number parity measurements, respectively, have been attempted. When photon on/off measurements 
with the perfect efficiency are used, the maximum value of the Bell violation is  ${\cal B}_{\cal O}\approx2.61$ at $|\alpha|\approx0.664$.
In order to see a violation of the Bell-CHSH inequality,
at least 50\% of detection efficiency is required.
When photon parity measurements are used, the value of the Bell-CHSH violation rapidly increases
as $\alpha$ gets larger, and it approaches Cirel'son's bound
 for $\alpha\gg1$. 
Although precise direct measurements of cavity
fields are experimentally difficult,
photon number parity measurements for the cavity field can be 
effectively performed 
using ancillary probe atoms and atomic detectors. 
We have fully analyzed decoherence effects in both field and atomic modes
and discuss conditions required to perform a Bell inequality test free from 
the locality loophole.

Our proposal may be considered an attempt to analyze a Bell inequality test
using entanglement between a microscopic system and a mesoscopic classical system.
Since atomic detectors are known to be highly efficient
\cite{Maioli2005}, it may also be a reasonable target to perform this type
of experiment in a way free from the detection loophole.
In principle, a Bell inequality test free from the locality loophole in our framework 
using atom-field entanglement 
may be performed using a long cylindrical cavity for the atom with a low spontaneous
emission rate  \cite{Kakazu1996}.
However, our analysis shows that it would be extremely demanding to perform a Bell inequality test
free from both the locality and detection loopholes in this framework since 
the main cavity for field with a low dissipation rate would be necessary together
with a long cylindrical cavity.

\section*{Acknowledgements}

J.P. and H.J. thank Chang-Woo Lee and Mauro Paternostro
for stimulating discussions.
This work was supported by the NRF grant funded by
the Korea government (MEST) (No. 3348-20100018)
and the World Class University (WCU) program.
J.P. acknowledges financial support from Seoul Scholarship Foundation,
and H.J. acknowledges support from TJ Park Foundation.

\appendix
\section{Solutions of the Master Equation for Matrix Elements}

We first find general solutions 
of the master equation (\ref{eq:mi})
for three types of decoherence processes step by step, {\it i.e.}, spontaneous emission of an atom, cavity dissipation, and atom-field interaction with spontaneous emission and cavity dissipation.

\subsection{Spontaneous emission for atom}
\label{subsection:Spontaneous Emission}

A density operator of a two-level atom, $\hat{\rho}_A (t)$, can be expressed as a matrix form
\begin{equation}
\hat \rho_A (t)  = \left( {\begin{array}{*{20}c}
   { \rho _{A,ee} (t)} & { \rho _{A,eg} (t)}  \\
   { \rho _{A,ge} (t)} & { \rho _{A,gg} (t)}  \\
\end{array}} \right),
\label{eq:rhoatom}
\end{equation} 
where 
${ \rho _{A,ij} (t)}=\left\langle i  \right| \hat \rho_A (t) \left| j  \right\rangle $.
When an atom with a initial density matrix, $\hat \rho_A (0)$, 
goes through the spontaneous emission process for time $t$, 
its density matrix is straightforwardly obtained using Eq.~(\ref{eq:mi}) with $\chi=0$ and $\kappa=0$ as
\begin{equation}
\begin{split}
\hat \rho_A (t)  =&{\hat {\cal S}}_A (\gamma, t) [\hat{\rho}_A (0) ] \\
=& \left( {\begin{array}{*{20}c}
   {e^{-2\gamma t} \rho _{A,ee} (0)} & {e^{-\gamma t} \rho _{A,eg} (0)}  \\
   {e^{-\gamma t} \rho _{A,ge} (0)} & { \rho _{A,gg} (0)- \rho _{A,ee} (0)(e^{-2\gamma t}-1)}  \\
\end{array}} \right),
\label{eq:rhogammati}
\end{split}
\end{equation} 
where superoperator $\hat {\cal S}(\gamma, t)$ is defined for later use.

\subsection{Dissipation for cavity field}
In order to find the time evolution of the coherent-state part the density operator, it is sufficient to find
the time evolution of an operator component 
$\left| \mu  \right\rangle \left\langle \nu  \right|$, where $\left| \mu  \right\rangle$ and $\left| \nu  \right\rangle$ are coherent states of amplitudes $\mu$ and $\nu$.
This solution 
for time $t$ under the master equation
(\ref{eq:mi}) with $\chi=0$ and $\gamma=0$ is well known as
\cite{Phoenix,Moya-Cessa2006Aug}
\begin{equation}
\exp[-\frac{({\left|\mu \right|}^2+{\left| \nu \right|}^2-2 \nu^* \mu)(1-\exp(-2\kappa t))}{2} ]\left| \mu e^{-\kappa t} \right\rangle \left\langle \nu  e^{-\kappa t} \right|.
\end{equation}

\subsection{Atom-field interaction with spontaneous emission and cavity dissipation}
\label{sec:HI}
The density matrix $\hat \rho (t)$ for an atom-field state can be considered in a
$2\times \infty$ dimensional space, since we assume a two-level atom.
It is possible to decompose the master equation (\ref{eq:mi}) in $\{ \left| e  \right\rangle,~\left| g  \right\rangle \}$ basis with the density matrix elements
${\hat \rho _{C,ij} (t)}=\left\langle i  \right| \hat \rho (t) \left| j  \right\rangle$.
We then obtain equations
\begin{equation}
\begin{split}
\frac{d}{{dt}}\hat \rho _{C,ee}  
=& {\hat {\cal{L}}}_{ee} \hat \rho _{C,ee}-2\gamma \hat \rho_{C,ee}\\
=&  - i\chi [\hat a^\dag  \hat a,\hat \rho _{C,ee} ] + \kappa (2\hat a\hat \rho _{C,ee} \hat a^\dag   - \hat a^\dag  \hat a\hat \rho _{C,ee}  - \hat \rho _{C,ee} \hat a^\dag  \hat a) \\
&- 2\gamma \hat \rho _{C,ee},\\
\end{split}
\label{eq:rhoIee}
\end{equation}
\begin{equation}
\begin{split}
\frac{d}{{dt}}\hat \rho _{C,gg}  
=& {\hat {\cal{L}}}_{gg} \hat \rho _{C,gg}+2\gamma \hat \rho_{C,ee}\\
=&   i\chi [\hat a^\dag  \hat a,\hat \rho _{C,gg} ] + 
\kappa (2\hat a\hat \rho _{C,gg} \hat a^\dag   - \hat a^\dag
  \hat a\hat \rho _{C,gg}  - \hat \rho _{C,gg} \hat a^\dag  \hat a) \\
  &+ 2\gamma \hat \rho _{C,ee},\\
\end{split}
\label{eq:rhoIgg}
\end{equation}
\begin{equation}
\begin{split}
\frac{d}{{dt}}\hat \rho _{C,eg}  
=& {\hat {\cal{L}}}_{eg} \hat \rho _{C,eg}-i\chi \hat \rho_{C,eg}-\gamma \hat \rho_{C,eg}\\
=&  - i\chi (\hat a^\dag  \hat a +1) \hat \rho _{C,eg} 
- i\chi \hat \rho _{C,eg}\hat a^\dag  \hat a \\
& + \kappa (2\hat a\hat \rho _{C,eg} \hat a^\dag   - \hat a^\dag  \hat a\hat \rho _{C,eg}  - \hat \rho _{C,eg} \hat a^\dag  \hat a) - \gamma \hat \rho _{C,eg},\\
\end{split}
\label{eq:rhoIeg}
\end{equation}
\begin{equation}
\begin{split}
\frac{d}{{dt}}\hat \rho _{C,ge}  
=& {\hat {\cal{L}}}_{ge} \hat \rho _{C,ge}+i\chi \hat \rho_{C,ge}-\gamma \hat \rho_{C,ge}\\
=&   i\chi \hat \rho _{C,ge}  (\hat a^\dag  \hat a +1) 
+ i\chi  \hat a^\dag  \hat a \hat \rho _{C,ge} \\
& + \kappa (2\hat a\hat \rho _{C,ge} \hat a^\dag   - \hat a^\dag
\hat a\hat \rho _{C,ge}  - \hat \rho _{C,ge} \hat a^\dag  \hat a) - \gamma \hat \rho _{C,ge}.
\end{split}
\label{eq:rhoIge}
\end{equation}
We define the following superoperators for simplicity:
${\hat {\cal{M}}}= \hat a^\dag  \hat a ~\cdot~$ ,
${\hat {\cal{P}}}=~\cdot~ \hat a^\dag  \hat a$  ,  and
${\hat {\cal{J}}}=\hat a  ~\cdot~  \hat a^\dag$
. Then ${\hat {\cal{L}}}_{ee}$, ${\hat {\cal{L}}}_{gg}$, ${\hat {\cal{L}}}_{eg}$, and ${\hat {\cal{L}}}_{ge}$ can be expressed as,
\begin{subequations}
\begin{align}
 {\hat {\cal{L}}}_{ee}  &\equiv 2\kappa {\hat {\cal{J}}} - r{\hat {\cal{M}}} - r^* {\hat {\cal{P}}}, \\ 
 {\hat {\cal{L}}}_{eg}  &\equiv 2\kappa {\hat {\cal{J}}} - r{\hat {\cal{M}}} - r{\hat {\cal{P}}},
 \end{align}
\end{subequations}
where $r\equiv \kappa+i\chi$, and 
${\hat {\cal{L}}}_{gg}$ and ${\hat {\cal{L}}}_{ge}$ are obtained by substituting $\chi$ with $-\chi$
in ${\hat {\cal{L}}}_{ee}$, and ${\hat {\cal{L}}}_{eg}$, respectively.
A master equation of the form
${d\hat \rho }/{dt} = \hat {\cal L} \hat \rho  + c\hat \rho$,
where $c$ is a constant and $\hat {\cal L}$ is a superoperator, can be solved with a usual exponential form 
$\exp{[(\hat {\cal L}+c) t]}\hat{\rho}$.

\subsubsection{Solution for ${\hat {\rho}}_{C,ee}$}
The solution of Eq.~(\ref{eq:rhoIee}) is
\begin{equation}
\hat \rho _{C,ee} (t) = \exp [(\hat {\cal L}_{ee}  - 2\gamma )t]\hat \rho _{C,ee} (0) = e^{ - 2\gamma t} e^{\hat {\cal L}_{ee} t}\hat \rho _{C,ee} (0).
\end{equation}
where the factorization can be done by the similarity transformation \cite{Witschel1981}.
Now we need to factorize $e^{\hat {\cal L}_{ee} t}$. This is solved with an ansatz (a technique can be found in 
Ref.~\cite{Moya-Cessa2006Aug})
\begin{equation}
\hat \rho _{C,ee} (t)= \exp[-2\gamma t]\exp[(- r{\hat {\cal{M}}} - r^* {\hat {\cal{P}}})t] \exp[f(t) 2\kappa {\hat {\cal{J}}}]\hat \rho_{C,ee}(0),
\end{equation}
where $f(t)=(1-e^{-2\kappa t})/(2\kappa)$.
For an initial state $\hat \rho_{C,ee}(0)=\left| \mu \right\rangle \left\langle \nu \right|$,  
\begin{equation}
\hat \rho _{C,ee} (t)=\exp[-2\gamma t + \Theta(\kappa, 0 ,\mu,\nu, t) ]
\left| \mu e^{-r t} \right\rangle \left\langle \nu  e^{-r t} \right|
\label{eq:rhoeemunu}
\end{equation}
with
\begin{equation}
\Theta(\kappa,\chi,\mu,\nu, t) := -\frac{1}{2} ({\left| \nu \right|}^2+{\left|\mu \right|}^2)(1-e^{-2\kappa t})  +\frac{\kappa}{r} (1-e^{-2r t}) \nu^* \mu.
\end{equation}

\subsubsection{Solution for ${\hat {\rho}}_{C,gg}$}
In order to solve Eq.~(\ref{eq:rhoIgg}), we first assume $\gamma=0$. A homogeneous solution is obtained from Eq.~(\ref{eq:rhoeemunu})
by substituting $\chi$ with $-\chi$ as
\begin{equation}
\hat \rho_{C,gg} ^h (t) = \exp[\Theta(\kappa,0,\mu,\nu, t) ]\left| \mu e^{-r^* t} \right\rangle \left\langle \nu  e^{-r^* t} \right|.
\end{equation}
Then it is obvious to see that the general solution $\hat \rho _{C,gg} (t)$ with $\gamma\ne0$ is
\begin{equation}
\begin{split}
\hat \rho_{C,gg} (t) &=\hat \rho_{C,gg} ^h (t)+2\gamma \int_0^t {dt' \hat \rho_{C,ee} (t')} \\
&=\exp[\Theta(\kappa,0,\mu,\nu, t) ]\left| \mu e^{-r^* t} \right\rangle \left\langle \nu  e^{-r^* t} \right| \\
 &+ 2\gamma \int_0^t {dt'\exp [ - 2\gamma t' + \Theta(\kappa,0,\mu,\nu, t')]} \left| {\mu e^{ - rt'} } 
 \right\rangle \left\langle {\nu e^{ - rt'} } \right|
.
\end{split}
\label{eq:rhoggmunu}
\end{equation}

\subsubsection{Solution for ${\hat {\rho}}_{C,ge}$}
The solution of Eq.~(\ref{eq:rhoIge}) is
\begin{equation}
\hat \rho _{C, ge} (t) = \exp [(\hat {\cal L}_{ge} +i \chi - \gamma )t]\hat \rho _{C,ge} (0) = e^{ (i\chi - \gamma) t} e^{\hat {\cal L}_{ge} t}\hat \rho _{C,ge} (0).
\end{equation}
Factoring $e^{\hat {\cal L}_{eg} t}$ with an ansatz 
\begin{equation}
\hat \rho _{C,ge} (t)= \exp[(i\chi-\gamma) t]\exp[(- r{\hat {\cal{M}}} - r^* {\hat {\cal{P}}})t] \exp[g(t) 2\kappa {\hat {\cal{J}}}]\hat \rho_{C,ee}(0),
\end{equation}
where $g(t)=(1-e^{-2r^* t})/(2r^*)$.
For $\hat \rho_{C,ge}(0)=\left| \mu \right\rangle \left\langle \nu \right|$,  
\begin{equation}
\hat \rho _{C,ge} (t)=\exp[(i\chi-\gamma) t +\Theta(\kappa,-\chi,\mu,\nu, t) ]
\left| \mu e^{-r^* t} \right\rangle \left\langle \nu  e^{-r t} \right|.
\label{eq:rhogemunu}
\end{equation}

\subsubsection{Solution for ${\hat {\rho}}_{C,eg}$}
The solution of Eq.~(\ref{eq:rhoIeg}) is obtained from Eq.~(\ref{eq:rhogemunu}) by substituting $\chi$
with $-\chi$  as
\begin{equation}
\hat \rho _{C,eg} (t)=\exp[(-i\chi-\gamma) t + \Theta(\kappa,\chi,\mu,\nu, t)  ]
\left| \mu e^{-r t} \right\rangle \left\langle \nu  e^{-r^* t} \right|.
\label{eq:rhoegmunu}
\end{equation}

\section{Derivation of the density matrices for atom-field entanglement and the correlation function}
\label{sec:DerivationofDM}
\subsection{Atom-field entanglement generated under decoherence effects}
First, atom $A$ initially prepared in
$| e \rangle_A$
 undergoes the spontaneous emission for the time $t_1$. 
After applying the first Ramsey pulse, $R_A=\hat{D}_A(-i\pi/4)$,
explained in Sec.~\ref{Chapter: Bell CHSH inequality with Wigner functions : Practical Scheme with Probing},
atom $A$ again undergoes the spontaneous emission for time $t_2$. Using Eqs.~(\ref{eq:rhogammati}) again,
it becomes
\begin{equation}
\begin{split}
&\hat{\cal S}_A(\gamma_0,t_2)\Big[
\hat{D}_A(-i\pi/4)|\Big\{
\hat{\cal S}_A(\gamma_0,t_1)[(|e\rangle\langle e|)_A]
\Big\}\hat{D}^\dagger_A(-i\pi/4)
\Big]\\
&~=\left( {\begin{array}{*{20}c}
   {\frac{1}{2}e^{ - 2\gamma_0 t_2 } } & {( - \frac{i}{2} + ie^{ - 2\gamma_0 t_1 } )e^{ - \gamma_0 t_2 } }  \\
   {(\frac{i}{2} - ie^{ - 2\gamma_0 t_1 } )e^{ - \gamma_0 t_2 } } & {(1 - \frac{1}{2}e^{ - 2\gamma_0 t_2 } )}  \\
\end{array}} \right).
\end{split}
\label{eq:atomAprepared}
\end{equation}
Then atom $A$ interacts with the cavity field $C$ prepared in state
$| i\alpha \rangle  _C$. 
Using Eqs.~(\ref{eq:rhoeemunu}), (\ref{eq:rhoggmunu}),  (\ref{eq:rhogemunu}) 
and (\ref{eq:rhoegmunu}),
we find the state
after the interaction time $t_3$ as $\hat{\rho}_{AC}^{(3)}$
with its matrix elements:
\begin{subequations}
\begin{equation}
\begin{split}
\hat \rho _{AC,ee} ^{(3)} = \frac{1}{2}e^{ - 2\gamma_0 t_2  - 2\gamma _c t_3 } \left| {i\alpha e^{ - rt_3 } } \right\rangle \left\langle {i\alpha e^{ - rt_3 } } \right|, 
\end{split}
\label{eq:initrhoee}
\end{equation}
\begin{equation}
\begin{split}
&\hat \rho _{AC,eg}^{(3)} = ( - \frac{i}{2} + ie^{ - 2\gamma_0 t_1 } ) \exp [ - \gamma_0 t_2+(- i\chi - \gamma _c)t_3   \\&+ \Theta(\kappa,\chi,\alpha,\alpha, t_3)]\left| {i\alpha e^{ - rt_3 } } \right\rangle \left\langle {i\alpha e^{ - r^* t_3 } } \right|,  
\end{split}
\label{eq:initrhoeg}
\end{equation}
\begin{equation}
\begin{split}
&\hat \rho _{AC,ge}^{(3)}  = (\frac{i}{2} - ie^{ - 2\gamma_0 t_1 } ) \exp [- \gamma_0 t_2+  ( i\chi - \gamma _c)t_3  \\
&+ \Theta(\kappa,-\chi,\alpha,\alpha, t_3)]\left| {i\alpha e^{ - r^* t_3 } } \right\rangle \left\langle {i\alpha e^{ - rt_3 } } \right|,
 \end{split}
 \label{eq:initrhoge}
\end{equation}
\begin{equation}
\begin{split}
&\hat \rho _{AC,gg} ^{(3)} =(1 - \frac{1}{2}e^{ - 2\gamma_0 t_2 } )\left| {i\alpha e^{ - r^* t_3 } } \right\rangle \left\langle {i\alpha e^{ - r^* t_3 } } \right| \\
&+ 2\gamma _c \int_0^{t_3 } {dt} \frac{1}{2}e^{ - 2\gamma_0 t_2  - 2\gamma _c t} \left| {i\alpha e^{ - rt} } \right\rangle \left\langle {i\alpha e^{ - rt} } \right|.
\end{split}
\label{eq:initrhogg}
\end{equation}
\end{subequations}

\subsection{Atom-field entanglement after traveling for the spacelike separation}
We now derive the total density matrix right before $\hat{\cal D}_C^{\dag}(\beta)$ is applied. The state $\hat \rho _{AC} ^{(3)}$ undergoes spontaneous emission inside the cylindrical cavity and dissipation inside the cavity field $C$. The calculation can be
done using the results in Sec.~\ref{sec:HI} with $\chi=0$. Then the state becomes $\hat{\rho}_{AC}^{(4)}$, where

\begin{subequations}
\begin{equation}
\begin{split}
\hat \rho _{AC,ee} ^{(4)}= \frac{1}{2}e^{ - 2\gamma_0 t_2  - 2\gamma _c t_3 -2\gamma' t_4} \left| {i\alpha e^{ - rt_3  -\kappa t_4} } \right\rangle \left\langle {i\alpha e^{ - rt_3 -\kappa t_4 } } \right|, 
\end{split}
\label{eq:BFDrhoee}
\end{equation}
\begin{equation}
\begin{split}
\hat \rho _{AC,eg} ^{(4)} &= ( - \frac{i}{2} + ie^{ - 2\gamma_0 t_1 } )\exp [ - \gamma_0 t_2+ (- i\chi - \gamma _c)t_3   - \gamma' t_4 \\&+\Theta(\kappa,\chi,\alpha,\alpha, t_3)+\Theta(\kappa,0,\alpha e^{-r t_3},\alpha e^{-r^* t_3}, t_4)] \\
&\left| {i\alpha e^{ - rt_3 - \kappa t_4 } } \right\rangle \left\langle {i\alpha e^{ - r^* t_3 - \kappa t_4} } \right|,  
\end{split}
\label{eq:BFDrhoeg}
\end{equation}
\begin{equation}
\begin{split}
\hat \rho _{AC,ge}^{(4)} &= (  \frac{i}{2} - ie^{ - 2\gamma_0 t_1 } )\exp [ - \gamma t_2+ ( i\chi - \gamma _c)t_3   - \gamma' t_4 \\&+\Theta(\kappa,-\chi,\alpha,\alpha, t_3)+\Theta(\kappa,0,\alpha e^{-r^* t_3},\alpha e^{-r t_3}, t_4)]\\
&\left| {i\alpha e^{ - r^* t_3 - \kappa t_4 } } \right\rangle \left\langle {i\alpha e^{ - r t_3 - \kappa t_4} } \right|,
\end{split}
 \label{eq:BFDrhoge}
\end{equation}
\begin{equation}
\begin{split}
\hat \rho _{AC,gg} ^{(4)} &=(1 - \frac{1}{2}e^{ - 2\gamma_0 t_2 } )\left| {i\alpha e^{ - r^* t_3 -\kappa t_4} } \right\rangle \left\langle {i\alpha e^{ - r^* t_3 -\kappa t_4} } \right| \\
&+ 2\gamma _c \int_0^{t_3 } {dt} \frac{1}{2}e^{ - 2\gamma_0 t_2  - 2\gamma _c t} \left| {i\alpha e^{ - rt -\kappa t_4} } \right\rangle \left\langle {i\alpha e^{ - rt -\kappa t_4} } \right|\\
&+ 2\gamma' \int_0^{t_4 } {dt} \frac{1}{2}e^{ - 2\gamma_0 t_2  - 2\gamma _c t_3 -2\gamma' t} \left| {i\alpha e^{ - rt_3 -\kappa t} } \right\rangle \left\langle {i\alpha e^{ - rt_3 -\kappa t} } \right|,
\end{split}
\label{eq:BFDrhogg}
\end{equation}
\end{subequations}
and here the subscripts are consistent with the previous ones.

\subsection{Effects with atom $B$ for indirect measurement}

After applying the displacement operation $\hat{\cal D}_C^{\dag} (\beta)$, the total state
 becomes  $\hat{\rho}_{AC}^{\beta}=\hat{\cal D}_C^{\dag} (\beta)\hat{\rho}_{AC}^l \hat{\cal D}_C (\beta)$.
Now, the probe atom $B$, which is in the same state as that of atom $A$
in Eq.~(\ref{eq:atomAprepared}), goes into the cavity field of state $\hat{\rho}_{AC}^{\beta}$. The atom-field interaction $H_I$ with the coupling constant $\chi$ occurs between atom $B$ and field $C$ for time $t_3$. When solving the master equation, it is convenient if one notes that the field-part of state $\hat{\rho}_{AC}^{\beta}$ can be expressed by coherent-state dyadics such as $|\mu \rangle\langle \nu|$. If the component of the cavity field, initially prepared as $|\mu \rangle\langle \nu|$, interacts with an atomic state (\ref{eq:atomAprepared}) for time $t_3$, the resulting density operator element is
obtained as $\hat{\Omega}_{BC}(\mu,\nu,t_3)$ with
\begin{subequations}
\begin{equation}
\begin{split}
\hat \Omega _{BC,ee} (\mu,\nu,t_3 ) = \frac{1}{2}e^{ - 2\gamma_0 t_2  - 2\gamma _c t_3 +\Theta (\kappa, 0,\mu,\nu, t_3)} \left| {\mu e^{ - rt_3 } } \right\rangle \left\langle {\nu e^{ - rt_3 } } \right|, 
\end{split}
\label{eq:omegaee}
\end{equation}
\begin{equation}
\begin{split}
&\hat \Omega _{BC,eg} ( \mu,\nu,t_3 ) = ( - \frac{i}{2} + ie^{ - 2\gamma_0 t_1 } ) \exp [ - \gamma_0 t_2+(- i\chi - \gamma _c)t_3   \\&+ \Theta(\kappa,\chi,\mu,\nu, t_3)]\left| {\mu e^{ - rt_3 } } \right\rangle \left\langle {\nu e^{ - r^* t_3 } } \right|,  
\end{split}
\label{eq:omegaeg}
\end{equation}
\begin{equation}
\begin{split}
&\hat \Omega _{BC,ge} (\mu,\nu,t_3 )  = (\frac{i}{2} - ie^{ - 2\gamma_0 t_1 } ) \exp [- \gamma_0 t_2+  ( i\chi - \gamma _c)t_3  \\
&+ \Theta(\kappa,-\chi,\mu,\nu, t_3)]\left| {\mu e^{ - r^* t_3 } } \right\rangle \left\langle {\nu e^{ - rt_3 } } \right|,
 \end{split}
 \label{eq:omegage}
\end{equation}
\begin{equation}
\begin{split}
&\hat \Omega _{BC,gg} (\mu,\nu, t_3 ) =(1 - \frac{1}{2}e^{ - 2\gamma_0 t_2 } )e^{   \Theta (\kappa,0,\mu,\nu,t_3)}\left| {\mu e^{ - r^* t_3 } } \right\rangle \left\langle {\nu e^{ - r^* t_3 } } \right| \\
&+ 2\gamma _c \int_0^{t_3 } {dt} \frac{1}{2}e^{ - 2\gamma_0 t_2  - 2\gamma _c t+ \Theta (\kappa,0,\mu,\nu,t)} \left| {\mu e^{ - rt} } \right\rangle \left\langle {\nu e^{ - rt} } \right|.
\end{split}
\label{eq:omegagg}
\end{equation}
\end{subequations}
We used  Eqs.~(\ref{eq:rhoeemunu}), (\ref{eq:rhoggmunu}),  (\ref{eq:rhogemunu}) and (\ref{eq:rhoegmunu})
again to find Eqs.~(\ref{eq:omegaee})-(\ref{eq:omegagg}).
Therefore, $\hat{\rho}_{AC}^{\beta}$ interacts with atom $B$ and evolves to
\begin{equation}
\begin{split}
\hat \rho_{ABC} 
=&( | e \rangle \langle e | )_A \otimes \hat {\rho}_{BC,ee}+
( | e \rangle \langle g | )_A \otimes \hat {\rho}_{BC,eg}\\
&+
( | g \rangle \langle e | )_A \otimes \hat {\rho}_{BC,ge}+
( | g \rangle \langle g | )_A \otimes \hat {\rho}_{BC,gg},
\end{split}
\end{equation}
where
\begin{subequations}
\begin{equation}
\begin{split}
\hat \rho _{BC,ee}
=&\frac{1}{2}e^{ - 2\gamma_0 t_2  - 2\gamma _c t_3 -2\gamma' (t_4+t_3)} 
\\
&\hat \Omega _{BC} ({i\alpha e^{ - rt_3  -\kappa t_4} -\beta},{i\alpha e^{ - rt_3 -\kappa t_4 } -\beta}, t_3 ), 
\end{split}
\label{eq:rhoABCee}
\end{equation}
\begin{equation}
\begin{split}
\hat \rho _{BC,eg}=& ( - \frac{i}{2} + ie^{ - 2\gamma_0 t_1 } )\exp [  - \gamma_0 t_2+ (- i\chi - \gamma _c)t_3   - \gamma' (t_4+t_3) \\
&+\Theta(\kappa,\chi,\alpha,\alpha, t_3)+\Theta(\kappa,0,\alpha e^{-r t_3},\alpha e^{-r^* t_3}, t_4)\\
&-2i e^{-\kappa(t_4+t_3)}\sin(\chi t_3){\rm{Im}} (\alpha^* \beta)]\\
&
\hat \Omega _{BC} ( {i\alpha e^{ - rt_3 - \kappa t_4 } -  \beta }, {i\alpha e^{ - r^* t_3 - \kappa t_4}  -  \beta} , t_3)  
\end{split}
\label{eq:rhoABCeg}
\end{equation}
\begin{equation}
\begin{split}
\hat \rho _{BC,ge}  =& (  \frac{i}{2} - ie^{ - 2\gamma_0 t_1 } )\exp [ - \gamma_0 t_2+ ( i\chi - \gamma _c)t_3   - \gamma' (t_4+t_3) \\&+\Theta(\kappa,-\chi,\alpha,\alpha, t_3)+\Theta(\kappa,0,\alpha e^{-r^* t_3},\alpha e^{-r t_3}, t_4) \\
&+2i e^{-\kappa(t_4+t_3)}\sin(\chi t_3){\rm{Im}} (\alpha^* \beta)] \\
&
\hat \Omega _{BC} ({i\alpha e^{ - r^* t_3 - \kappa t_4} -  \beta } , {i\alpha e^{ - r t_3 - \kappa t_4} -  \beta} , t_3)
\end{split}
 \label{eq:rhoABCge}
\end{equation}
\begin{equation}
\begin{split}
&\hat \rho _{BC,gg}
=(1 - \frac{1}{2}e^{ - 2\gamma_0 t_2 } )\hat \Omega _{BC} ( {i\alpha e^{ - r^* t_3 -\kappa t_4} -\beta} , {i\alpha e^{ - r^* t_3 -\kappa t_4} -\beta} ,t_3) \\
&+ 2\gamma _c \int_0^{t_3 } {dt} \frac{1}{2}e^{ - 2\gamma_0 t_2  - 2\gamma _c t} \hat \Omega _{BC}( {i\alpha e^{ - rt -\kappa t_4} -\beta} , {i\alpha e^{ - rt -\kappa t_4} -\beta} ,t_3)\\
&+ 2\gamma' \int_0^{t_4 } {dt} \frac{1}{2}e^{ - 2\gamma_0 t_2  - 2\gamma _c t_3 -2\gamma' t} \hat \Omega _{BC}( {i\alpha e^{ - rt_3 -\kappa t} -\beta}, {i\alpha e^{ - rt_3 -\kappa t} -\beta} ,t_3)\\
&+\frac{1}{2}e^{ - 2\gamma_0 t_2  - 2\gamma _c t_3 -2\gamma' t_4} 
\hat \Omega _{BC} ({i\alpha e^{ - rt_3  -\kappa t_4} -\beta},{i\alpha e^{ - rt_3 -\kappa t_4 } -\beta}, t_3 )\\
&\times(1-e^{-2\gamma' t_3}).
\end{split}
\label{eq:rhoABCgg}
\end{equation}
\end{subequations}
Spontaneous emissions of atom $A$ and $B$ that may occur after this point shall be taken into account when we derive the correlation function. Since the field state is not considered any more from this point before the final measurements,
the cavity dissipation can be ignored. By tracing out the cavity field, we get
\begin{equation}
\begin{split}
\hat \rho _{AB}={\rm Tr}_C \hat \rho _{ABC}
=\sum\limits_{i,j=e,g} {( | i \rangle \langle j | )_A \otimes \hat {\sigma}_{B,ij}},
\label{eq:rhoAB}
\end{split}
\end{equation}
where
\begin{subequations}
\begin{equation}
\begin{split}
{\hat \sigma}_{B,ee}
=&{\rm{Tr}}_C \hat {\rho}_{BC,ee}\\
=&\frac{1}{2}e^{ - 2\gamma_0 t_2  - 2\gamma _c t_3 -2\gamma' (t_4+t_3)} \\
&\hat \mho _{B} ({i\alpha e^{ - rt_3  -\kappa t_4} -\beta},{i\alpha e^{ - rt_3 -\kappa t_4 } -\beta}, t_3 ), 
\end{split}
\label{eq:rhoABee}
\end{equation}
\begin{equation}
\begin{split}
{\hat \sigma}_{B,eg}=& ( - \frac{i}{2} + ie^{ - 2\gamma_0 t_1 } )\exp [  - \gamma_0 t_2+ (- i\chi - \gamma _c)t_3   - \gamma' (t_4+t_3) \\
&+\Theta(\kappa,\chi,\alpha,\alpha, t_3)+\Theta(\kappa,0,\alpha e^{-r t_3},\alpha e^{-r^* t_3}, t_4)\\
&-2i e^{-\kappa(t_4+t_3)}\sin(\chi t_3){\rm{Im}} (\alpha^* \beta)]\\
&
\hat \mho _{B} ( {i\alpha e^{ - rt_3 - \kappa t_4 } -  \beta }, {i\alpha e^{ - r^* t_3 - \kappa t_4}  -  \beta} , t_3),
\end{split}
\label{eq:rhoABeg}
\end{equation}
\begin{equation}
\begin{split}
{\hat \sigma}_{B,ge}  =& (  \frac{i}{2} - ie^{ - 2\gamma_0 t_1 } )\exp [ - \gamma_0 t_2+ ( i\chi - \gamma _c)t_3   - \gamma' (t_4+t_3) \\&+\Theta(\kappa,-\chi,\alpha,\alpha, t_3)+\Theta(\kappa,0,\alpha e^{-r^* t_3},\alpha e^{-r t_3}, t_4) \\
&+2i e^{-\kappa(t_4+t_3)}\sin(\chi t_3){\rm{Im}} (\alpha^* \beta)] \\
&
\hat \mho _{B} ({i\alpha e^{ - r^* t_3 - \kappa t_4} -  \beta } , {i\alpha e^{ - r t_3 - \kappa t_4} -  \beta} , t_3),
\end{split}
 \label{eq:rhoABge}
\end{equation}
\begin{equation}
\begin{split}
&{\hat \sigma}_{B,gg}
=(1 - \frac{1}{2}e^{ - 2\gamma_0 t_2 } )\hat \mho _{B} ( {i\alpha e^{ - r^* t_3 -\kappa t_4} -\beta} , {i\alpha e^{ - r^* t_3 -\kappa t_4} -\beta} ,t_3) \\
&+ 2\gamma _c \int_0^{t_3 } {dt} \frac{1}{2}e^{ - 2\gamma_0 t_2  - 2\gamma _c t} \hat \mho _{B}( {i\alpha e^{ - rt -\kappa t_4} -\beta} , {i\alpha e^{ - rt -\kappa t_4} -\beta} ,t_3)\\
&+ 2\gamma' \int_0^{t_4 } {dt} \frac{1}{2}e^{ - 2\gamma_0 t_2  - 2\gamma _c t_3 -2\gamma' t} \hat \mho _{B}( {i\alpha e^{ - rt_3 -\kappa t} -\beta}, {i\alpha e^{ - rt_3 -\kappa t} -\beta} ,t_3)\\
&+\frac{1}{2}e^{ - 2\gamma_0 t_2  - 2\gamma _c t_3 -2\gamma' t_4} 
\hat \mho _{B} ({i\alpha e^{ - rt_3  -\kappa t_4} -\beta},{i\alpha e^{ - rt_3 -\kappa t_4 } -\beta}, t_3 )\\
&\times(1-e^{-2\gamma' t_3}),
\end{split}
\label{eq:rhoABgg}
\end{equation}
\end{subequations}
and operator $\hat \mho _B (\mu, \nu, t_3)= {\rm Tr}_C \hat \Omega_{BC}(\mu,\nu,t_3)$ is determined as
\begin{subequations}
\begin{equation}
\begin{split}
 \mho _{B,ee} (\mu,\nu,t_3 ) = \frac{1}{2}e^{ - 2\gamma_0 t_2  - 2\gamma _c t_3+\Theta(\kappa,0,\mu,\nu,t_3) -\frac{1}{2}
(|\mu|^2+|\nu|^2-2\mu\nu^*) \exp(-2\kappa t_3)} 
, 
\end{split}
\label{eq:mhoee}
\end{equation}
\begin{equation}
\begin{split}
&\mho _{B,eg} ( \mu,\nu,t_3 ) = ( - \frac{i}{2} + ie^{ - 2\gamma_0 t_1 } ) \exp [ - \gamma_0 t_2+(- i\chi - \gamma _c)t_3   \\&+ \Theta(\kappa,\chi,\mu,\nu, t_3)-\frac{1}{2}
(|\mu|^2+|\nu|^2-2\mu\nu^* e^{-2i \chi t_3}) e^{-2\kappa t_3}],
\end{split}
\label{eq:mhoeg}
\end{equation}
\begin{equation}
\begin{split}
&\mho _{B,ge} (\mu,\nu,t_3 )  = (\frac{i}{2} - ie^{ - 2\gamma_0 t_1 } ) \exp [- \gamma_0 t_2+  ( i\chi - \gamma _c)t_3  \\
&+ \Theta(\kappa,-\chi,\mu,\nu, t_3)-\frac{1}{2}
(|\mu|^2+|\nu|^2-2\mu\nu^* e^{2i \chi t_3}) e^{-2\kappa t_3}],
 \end{split}
 \label{eq:mhoge}
\end{equation}
\begin{equation}
\begin{split}
&\mho _{B,gg} (\mu,\nu, t_3 ) =(1 - \frac{1}{2}e^{ - 2\gamma t_2 } )e^{ \Theta(\kappa,0,\mu,\nu,t_3)-\frac{1}{2}
(|\mu|^2+|\nu|^2-2\mu\nu^*) \exp(-2\kappa t_3)} \\
&+ 2\gamma _c \int_0^{t_3 } {dt} \frac{1}{2}e^{ - 2\gamma_0 t_2  - 2\gamma _c t +\Theta(\kappa,0,\mu,\nu,t)-\frac{1}{2}
(|\mu|^2+|\nu|^2-2\mu\nu^*) \exp(-2\kappa t)} .
\end{split}
\label{eq:mhogg}
\end{equation}
\end{subequations}

\subsection{Decoherence right before final measurements and the correlation function}
We now consider the last measurement process for both parties. Atom $A$ experiences spontaneous emission for time $t_5-t_3$ with rate $\gamma^{\prime}$, then atomic displacement operation $\hat{D}^{\dag}_A(-e^{-i \phi}\pi/4)$ is applied. After the displacement operation, atom $A$ evolves again under the spontaneous emission for time $t_1$ with rate $\gamma_0$. We define superoperator ${\hat {\cal X}}$ to describe this process as
\begin{equation}
\begin{split}
&{\hat {\cal X}}_A(\gamma^{\prime},\gamma_0,t_5-t_3,t_1,\phi)\big[\hat \rho_A\big]\\
&={\hat {\cal S}}_A(\gamma_0,t_1)
\Big[\hat{D}^{\dag}_A(-e^{-i \phi}\pi/4)
\Big\{
{\hat {\cal S}}_A(\gamma^\prime, t_5-t_3)\big[\hat \rho_A\big]\Big\}\hat{D}_A(-e^{-i \phi}\pi/4)
\Big]
\end{split}
\label{eq:superop}
\end{equation}
Atom $B$ undergoes spontaneous emission for time $t_2$ with rate $\gamma_0$, and displacement operation $\hat{D}^{\dag}_B(-\pi/4)$ is applied. Then, it experiences spontaneous emission for time $t_1$ with rate $\gamma_0$ just before the final measurement.
This process can be expressed as
\begin{equation}
\begin{split}
&{\hat {\cal X}}_B(\gamma_0,\gamma_0,t_2,t_1,0)\big[\hat \rho_B\big]\\
&={\hat {\cal S}}_B(\gamma_0,t_1)\Big[\hat{D}^{\dag}_B(-\pi/4)
\Big\{{\hat {\cal S}}_B(\gamma_0, t_2)\big[\hat \rho_B\big]\Big\} \hat{D}_B(-\pi/4)\Big]
\end{split}
\end{equation}
The final density operator used to obtain the correlation function is 
then obtained using  state $\hat {\rho}_{AB}$ in Eq.~(\ref{eq:rhoAB}) 
with ${\hat {\cal X}}_A$ and ${\hat {\cal X}}_B$ as 
\begin{equation}
\hat\rho_{AB}^{\rm final}={\hat {\cal X}}_A(\gamma^{\prime},\gamma_0,t_5-t_3,t_1,\phi) \otimes {\hat {\cal X}}_B(\gamma_0,\gamma_0,t_2,t_1,0)\big[\hat\rho_{AB}\big].
\end{equation}
The correlation function is obtained as
the expectation value of dichotomic measurements (\ref{eq:dchm}) performed by both the parties:
\begin{widetext}
\begin{equation}
\begin{split}
E&(\phi,\beta,t_1,t_2,t_3,t_4,t_5)={\rm Tr} [\hat \rho_{AB}^{\rm final} \hat \Gamma_A \otimes \hat \Gamma_B]\\
=&\frac{1}{2}e^{ - 2\gamma_0 t_2  - 2\gamma _c t_3 -2\gamma' t_4}(e^{-2\gamma_0 t_1}-1)(1-e^{-2\gamma' t_3}+e^{-2\gamma' t_5})
\xi _{B} (\Lambda_1,\Lambda_1, t_3 ) + {\cal Z}~ \xi _{B} ( \Lambda_1, \Lambda_2 , t_3)  + {\cal Z}^*
\xi _{B} (\Lambda_2, \Lambda_1 , t_3)\\
&
+(e^{-2\gamma_0 t_1}-1)\{(1 - \frac{1}{2}e^{ - 2\gamma_0 t_2 } )
 \xi _{B} ( \Lambda_2 , \Lambda_2 ,t_3) + 2\gamma _c \int_0^{t_3 } {dt} \frac{1}{2}e^{ - 2\gamma_0 t_2  - 2\gamma _c t}  \xi_{B}( {i\alpha e^{ - rt -\kappa t_4} -\beta} , {i\alpha e^{ - rt -\kappa t_4} -\beta} ,t_3)\\
&+ 2\gamma' \int_0^{t_4 } {dt} \frac{1}{2}e^{ - 2\gamma_0 t_2  - 2\gamma _c t_3 -2\gamma' t}  \xi _{B}( {i\alpha e^{ - rt_3 -\kappa t} -\beta}, {i\alpha e^{ - rt_3 -\kappa t} -\beta} ,t_3)\},
\end{split}
\label{eq:correlationfunction}
\end{equation}
where
\begin{equation}
\begin{split}
&\xi_B(\mu,\nu,t_3)=( \mho_{B,ee}(\mu,\nu,t_3)+ \mho_{B,gg}(\mu,\nu,t_3))(e^{-2 \gamma_0 t_1}-1)+(\mho_{B,eg}(\mu,\nu,t_3)+\mho_{B,ge}(\mu,\nu,t_3))e^{-\gamma_0 t_5-2 \gamma_0 t_1},\\
&{\cal Z} =( - \frac{i}{2} + ie^{ - 2\gamma_0 t_1 } )e^{   - \gamma_0 (t_2+2 t_1)+ (- i\chi - \gamma _c)t_3   - \gamma' (t_4+t_5) +\Theta(\kappa,\chi,\alpha,\alpha, t_3)+\Theta(\kappa,0,\alpha e^{-r t_3},\alpha e^{-r^* t_3}, t_4)
-2i e^{-\kappa(t_4+t_3)}\sin(\chi t_3){\rm{Im}} (\alpha^* \beta)+i \phi},
\end{split}
\end{equation}
\end{widetext}
$\Lambda_1={i\alpha e^{ - rt_3  -\kappa t_4} -\beta}$ and $\Lambda_2={i\alpha e^{ - r^*t_3  -\kappa t_4} -\beta}$.
Using this correlation function, one can eventually construct the Bell function using Eq.~(\ref{eq:Bell CHSH}).


\begin{thebibliography}{57}
\bibitem{Einstein1935}%
A. Einstein, B. Podolsky, and N. Rosen, Phys. Rev. \textbf{47}, 777 (1935).
\bibitem{Bell1964}%
J. S. Bell, Physics \textbf{1}, 195 (1964).
\bibitem{Clauser1969}%
J. F. Clauser, M. A. Horne, A. Shimony, and R. A. Holt, Phys.
Rev. Lett. \textbf{23}, 880 (1969).
\bibitem{Freedman1972}%
S. J. Freedman and J. F. Clauser, Phys. Rev. Lett. \textbf{28}, 938
(1972).
\bibitem{Aspect1981}%
A. Aspect, P. Grangier, and G. Roger, Phys. Rev. Lett. \textbf{47}, 460
(1981).
\bibitem{Tittel1998}%
W. Tittel, J. Brendel, H. Zbinden, and N. Gisin, Phys. Rev. Lett.
\textbf{81}, 3563 (1998).
\bibitem{Weihs1998}%
G. Weihs, T. Jennewein, C. Simon, H. Weinfurter, and
A. Zeilinger, Phys. Rev. Lett. \textbf{81}, 5039 (1998).
\bibitem{Pearle1970}%
P. M. Pearle, Phys. Rev. D \textbf{2}, 1418 (1970).
\bibitem{Rowe2001}%
M. A. Rowe, D. Kielpinski, V. Meyer, C. A. Sackett, W. M.
Itano, C. Monroe, and D. J. Wineland, Nature \textbf{409}, 791 (2001).
\bibitem{Matsukevich2008}%
D. N. Matsukevich, P. Maunz, D. L. Moehring, S. Olmschenk,
and C. Monroe, Phys. Rev. Lett. \textbf{100}, 150404 (2008).
\bibitem{Bell1981}%
J. S. Bell, J. Phys. C \textbf{2}, 41 (1981).
\bibitem{MSKim2000}%
M. S. Kim and J. Lee, Phys. Rev. A \textbf{61}, 042102 (2000).
\bibitem{Milman2005}%
P. Milman, A. Auffeves, F. Yamaguchi, M. Brune, J. M. Rai-
mond, and S. Haroche, Eur. Phys. J. D \textbf{32}, 233 (2005).
\bibitem {Simon2003}%
C. Simon and W. T. M. Irvine, Phys. Rev. Lett. \textbf{91}, 110405
(2003).
\bibitem{Volz2006}%
J. Volz, M. Weber, D. Schlenk, W. Rosenfeld, J. Vrana,
K. Saucke, C. Kurtsiefer, and H. Weinfurter, Phys. Rev. Lett. \textbf{96}, 030404 (2006).
\bibitem{Brunner2007}
N. Brunner, N. Gisin, V. Scarani, and C. Simon, Phys. Rev.
Lett. \textbf{98}, 220403 (2007).
\bibitem{San2011} 
N. Sangouard, J.-D. Bancal, N. Gisin, W. Rosenfeld, P. Sekatski, M. Weber, and H. Weinfurter, Phys. Rev. A \textbf{84},
052122 (2011).
\bibitem{Spa2011}
N. Spagnolo, C. Vitelli, M. Paternostro, F. De Martini, and
F. Sciarrino, Phys. Rev. A \textbf{84}, 032102 (2011).
\bibitem{Moehring2004}%
D. L. Moehring, M. J. Madsen, B. B. Blinov, and C. Monroe,
Phys. Rev. Lett. \textbf{93}, 090410 (2004).
\bibitem{Wodkiewicz2000}%
 K. W\'{o}dkiewicz, New J. Phys. \textbf{2}, 21 (2000).
\bibitem{Schrodinger1935}
 E. Schr{\"o}dinger, Naturwissenschaften \textbf{23}, 823 (1935).%
\bibitem{Brune1996}%
M. Brune, E. Hagley, J. Dreyer, X. Ma\^{i}tre, A. Maali, C. Wunderlich, J. M. Raimond, and S. Haroche, Phys. Rev. Lett. \textbf{77},
4887 (1996).
\bibitem{Guerlin2007}%
C. Guerlin, J. Bernu, S. Del\'{e}glise, C. Sayrin, S. Gleyzes,
S. Kuhr, M. Brune, J.-M. Raimond, and S. Haroche, Nature
\textbf{448}, 889 (2007).
\bibitem{Deleglise2008}%
S. Del\'{e}glise, I. Dotsenko, C. Sayrin, J. Bernu, M. Brune, J.-M. Raimond, and S. Haroche, Nature \textbf{455}, 510 (2008).
\bibitem{Jeong2006Sep}%
H. Jeong and T. C. Ralph, Phys. Rev. Lett. \textbf{97}, 100401 (2006);
Phys. Rev. A 76, 042103 (2007).
\bibitem{Martini2008Jun}%
F. De Martini, F. Sciarrino, and C. Vitelli, Phys. Rev. Lett. \textbf{100},
253601 (2008).%
\bibitem{Spagnolo2010Nov}%
N. Spagnolo, C. Vitelli, F. Sciarrino, and F. De Martini, Phys.
Rev. A \textbf{82}, 052101 (2010).
\bibitem{Munro2000}
W. J. Munro, G. J. Milburn, and B. C. Sanders, Phys. Rev. A \textbf{62}, 052108 (2000).
\bibitem{Wilson2002}%
D. Wilson, H. Jeong, and M. S. Kim, J. Mod. Opt., Special
Issue for QEP \textbf{15}, 851 (2002).
\bibitem{Jeong2003}%
H. Jeong, W. Son, M. S. Kim, D. Ahn, and {\v C}. Brukner, Phys.
Rev. A \textbf{67}, 012106 (2003).
\bibitem{JeongAn2006}%
H. Jeong and N. B. An, Phys. Rev. A \textbf{74}, 022104 (2006).
\bibitem{Stobinska2007}%
M. Stobi\ifmmode \acute{n}\else \'{n}\fi{}ska, H. Jeong, and T. C. Ralph, Phys. Rev. A \textbf{75},
052105 (2007).
\bibitem {JeongSole2008}%
H. Jeong, Phys. Rev. A \textbf{78}, 042101 (2008).
\bibitem{Jeong2009}%
H. Jeong, M. Paternostro, and T. C. Ralph, Phys. Rev. Lett. \textbf{102}, 060403 (2009).
\bibitem{Gerry2009}
C. C. Gerry, A. Benmoussa, E. E. Hach, and J. Albert, Phys.
Rev. A \textbf{79}, 022111 (2009).
\bibitem{LeeJeong2009} C.-W. Lee and H. Jeong, Phys. Rev. A {\bf 80}, 052105 (2009).
\bibitem{Cirelson1980}%
B. S. CirelÕson, Lett. Math. Phys. \textbf{4}, 93 (1980).
\bibitem {Haroche2006}%
S. Haroche and J. Raimond, \textit{Exploring the Quantum, Atoms, Cavities, and Photons} (Oxford University Press, New York, 2006).
\bibitem{Yuen1980}%
H. Yuen and J. Shapiro, IEEE Trans. Inf. Theory \textbf{26}, 78 (1980).
\bibitem{Campos1989}%
R. A. Campos, B. E. A. Saleh, and M. C. Teich, Phys. Rev. A
\textbf{40}, 1371 (1989).
\bibitem {Pres1988}  
W. H. Press, B. P. Flannery, S. A. Teukolsky, and W. T. Vetterling, \textit{Numerical Recipes} (Cambridge   University 
Press, Cambridge, 1988).
\bibitem{Gisin1999}
N. Gisin and B. Gisin, Phys. Lett. A \textbf{260}, 323 (1999).
\bibitem{Brune1992}
 M. Brune, S. Haroche, J. M. Raimond, L. Davidovich, and
N. Zagury, Phys. Rev. A \textbf{45}, 5193 (1992).
\bibitem {Davidovich1996}%
L. Davidovich, M. Brune, J. M. Raimond, and S. Haroche, Phys. Rev. A \textbf{53}, 1295 (1996).
\bibitem {Raimond2001}%
J. M. Raimond, M. Brune, and S. Haroche, Rev. Mod. Phys. \textbf{73}, 565 (2001).
\bibitem {Nussenzveig1993}
P. Nussenzveig, F. Bernardot, M. Brune, J. Hare, J. M. Raimond, S. Haroche, and W. Gawlik, Phys. Rev. A \textbf{48}, 3991
(1993).
\bibitem{Englert1993}%
B.-G. Englert, N. Sterpi, and H. Walther, Optics Communications \textbf{100}, 526 (1993).
\bibitem{Hulet1985}%
R. G. Hulet, E. S. Hilfer, and D. Kleppner, Phys. Rev. Lett. \textbf{55},
2137 (1985).
\bibitem{Kakazu1996}%
K. Kakazu and Y. S. Kim, Prog. Theor. Phys. \textbf{96}, 883 (1996).
\bibitem{Wilkens1992}%
 M. Wilkens, Z. Bialynicka-Birula, and P. Meystre, Phys. Rev.
A \textbf{45}, 477 (1992).
\bibitem{Faria1999May}%
J. G. Peixoto de Faria and M. C. Nemes, Phys. Rev. A \textbf{59}, 3918
(1999).
\bibitem{Zhou2009P}
X. Zhou, C. Sayrin, S. Del\'{e}glise, J. Bernu, C. Guerlin, S. Gleyzes, S. Kuhr, I. Dotsenko, J.-M. Raimond, and S. Haroche, ``http://www.cqed.org/img/pdf/2009-lkb-aeres-techniqueslow.pdf''.
\bibitem{Kuhr2007}
S. Kuhr, S. Gleyzes, C. Guerlin, J. Bernu, U. B. Hoff, S. Del\'{e}glise, S. Osnaghi, M. Brune, J.-M. Raimond, S. Haroche, E. Jacques, P. Bosland, and B. Visentin, Appl. Phys. Lett. \textbf{90}, 164101 (2007).
\bibitem{Maioli2005}%
P. Maioli, T. Meunier, S. Gleyzes, A. Auffeves, G. Nogues, M. Brune, J. M. Raimond, and S. Haroche, Phys. Rev. Lett. \textbf{94}, 113601 (2005).
\bibitem{Phoenix}
S. J. D. Phoenix, Phys. Rev. A \textbf{41}, 5132 (1990).
\bibitem{Moya-Cessa2006Aug}
H. Moya-Cessa, Phys. Rep. \textbf{432}, 1 (2006).
\bibitem{Witschel1981}
W. Witschel, Int. J. Quantum Chem. \textbf{20}, 1233 (1981).
\end{thebibliography}
\end{document}